\documentclass[twocolumn, pra, showpacs, preprintnumbers, amsmath, amssymb, superscriptaddress]{revtex4}

\usepackage[pdftex]{graphicx}
\usepackage{bm}
\usepackage{subfigure}

\begin{document}

\newcommand{\kt}{\gamma}
\newcommand{\lzt}{q_z}
\newcommand{\atled}{\bm{\nabla}}
\newcommand{\dx}{\frac{\partial}{\partial_x}}
\newcommand{\dy}{\frac{\partial}{\partial_y}}
\newcommand{\dz}{\frac{\partial}{\partial_z}}
\newcommand{\dt}{\frac{\partial}{\partial_t}}
\newcommand{\sqrdt}{\frac{\partial^2}{\partial_t^2}}
\newcommand{\pbyp}[2]{\frac{\partial #1}{\partial #2}}
\newcommand{\dbyd}[2]{\frac{d #1}{d #2}}
\newcommand{\ex}{\bm{e}_x}
\newcommand{\ey}{\bm{e}_y}
\newcommand{\ez}{\bm{e}_z}
\newcommand{\besselj}[2]{\mathrm{J}_{#1}(#2)}
\newcommand{\besseljp}[2]{\mathrm{J'}_{#1}(#2)}
\newcommand{\hankel}[3]{\mathrm{H}_{#1}^{(#2)}(#3)}
\newcommand{\hankelp}[3]{\mathrm{H'}_{#1}^{(#2)}(#3)}
\newcommand{\laplace}{\Delta}
\newcommand{\neff}{n_{\mathrm{eff}}}
\newcommand{\fexp}{f_{\mathrm{expt}}}
\newcommand{\ftheo}{f_{\mathrm{calc}}}
\newcommand{\nexp}{\tilde{n}}
\newcommand{\Gtheo}{\Gamma_{\mathrm{calc}}}
\newcommand{\Gexp}{\Gamma_{\mathrm{expt}}}
\newcommand{\Grad}{\Gamma_{\mathrm{rad}}}
\newcommand{\Gabs}{\Gamma_{\mathrm{abs}}}
\newcommand{\Gant}{\Gamma_{\mathrm{ant}}}
\newcommand{\reffig}[1]{\mbox{Fig. \ref{#1}}}
\newcommand{\subreffig}[1]{\mbox{Fig. \subref{#1}}}
\newcommand{\refeq}[1]{\mbox{Eq. (\ref{#1})}}
\renewcommand{\Re}[1]{\mathrm{Re}(#1)}
\renewcommand{\Im}[1]{\mathrm{Im}(#1)}

\hyphenation{re-so-nan-ce re-so-nan-ces ex-ci-ta-tion z-ex-ci-ta-tion di-elec-tric ap-pro-xi-ma-tion ra-dia-tion}


\title{Experimental Test of a Two-dimensional Approximation for Dielectric Microcavities}

\author{S. Bittner}
\author{B. Dietz}
\author{M. Miski-Oglu}
\author{P. Oria Iriarte}
\affiliation{Institut f\"ur Kernphysik, Technische Universit\"at Darmstadt, D-64289 Darmstadt, Germany}
\author{A. Richter}
\email{richter@ikp.tu-darmstadt.de}
\affiliation{Institut f\"ur Kernphysik, Technische Universit\"at Darmstadt, D-64289 Darmstadt, Germany}
\affiliation{ECT*, Villa Tambosi, I-38050 Villazano (Trento), Italy}
\author{F. Sch\"afer}
\affiliation{Institut f\"ur Kernphysik, Technische Universit\"at Darmstadt, D-64289 Darmstadt, Germany}

\date{\today}

\begin{abstract}
Open dielectric resonators of different shapes are widely used for the manufacture of microlasers. A precise determination of their resonance frequencies and widths is crucial for their design. Most microlasers have a flat cylindrical geometry, and a two-dimensional approximation, the so-called method of the effective index of refraction, is commonly employed for numerical calculations. Our aim has been an experimental test of the precision and applicability of a model based on this approximation. We performed very thorough and accurate measurements of the resonance frequencies and widths of two passive circular dielectric microwave resonators and found significant deviations from the model predictions. From this we conclude that the model generally fails in the quantitative description of three-dimensional dielectric resonators.
\end{abstract}

\pacs{42.55.Sa, 05.45.Mt, 42.60.Da}

\maketitle

\section{\label{sec:intr}Introduction}
Open dielectric resonators are used in a large variety of applications, ranging from radio frequency and millimeter-wave applications \cite{Annino1997, Annino2000} to microlasers \cite{Noeckel2002, McCall1992, Vahala2004}. Therefore, accurate model predictions for their spectra and field distributions are of great interest. Especially microlasers have received much attention lately in optical telecommunication, as sensors or as billiard models \cite{Vahala2004, Krioukov2002, Armani2006, Annino2000}. These devices typically consist of a flat cylindrical dielectric resonator with a cross section of arbitrary shape which contains the active medium. The resonators are usually made of semiconductor \cite{McCall1992, Chu1993, Frateschi1996} or organic materials \cite{Kuwata-Gonokami1995, Polson2004, Lebental2006} and are sandwiched between two media of lower index of refraction like air or a substrate. The exact shape of the resonator is important in view of the applications, because it determines the emission properties like the directionality of radiation and the quality factors of the resonances \cite{Noeckel1994, Noeckel1997, Gmachl1998}. \newline
In general, even simple geometries like a flat dielectric disk with a height much smaller than the planar extension cannot be solved analytically. Since the numerical solution of the three-dimensional (3D) vectorial Maxwell equations describing such resonators is complicated and computationally demanding, suitable approximations are favorable. One widely used approximation is the reduction of the full 3D Maxwell equations to a two-dimensional (2D), scalar Helmholtz equation by introducing a so-called effective index of refraction $\neff$ (see e.g.\ \cite{Chin1994, Casey1978, Lebental2007, Dubertrand2008}). This 2D-approximation seems natural due to the flat shape with large extension in the plane of the microlasers, but to our knowledge its validity and precision has never been rigorously tested. The aim of the work presented here is thus the comparison of the experimentally measured resonance frequencies and widths with those calculated using this 2D-approximation, called the $\neff$-model in the following. \newline
Although motivated by microlasers working in the infrared to the optical spectrum, our experimental setup consists of a flat cylindrical microwave resonator made of Teflon (from the company Gr\"unberg Kunststoffe GmbH). Microwave experiments have distinct experimental advantages, and therefore are commonly used for the investigation of 2D quantum billiards \cite{richter97, StoeckmannBuch2000} and also 2D dielectric resonators \cite{Schaefer2006}. The results from the microwave resonators can, however, be directly applied to microcavities by scaling. The plan of the present article is the following. In section \ref{sec:neff}, the concept of an effective index of refraction is introduced and section \ref{sec:dielcircle} explains how dielectric resonators are modeled with it. The experimental setup is detailed in section \ref{sec:expsetup}, and the experimental data are compared to the model calculations for two disks of different thickness in sections \ref{sec:HKreis} and \ref{sec:NGKreis}. Finally, the results are discussed in section \ref{sec:conc}.

\section{\label{sec:neff}Effective Index of Refraction}

\begin{figure}[b]
\includegraphics[width = 8.6 cm]{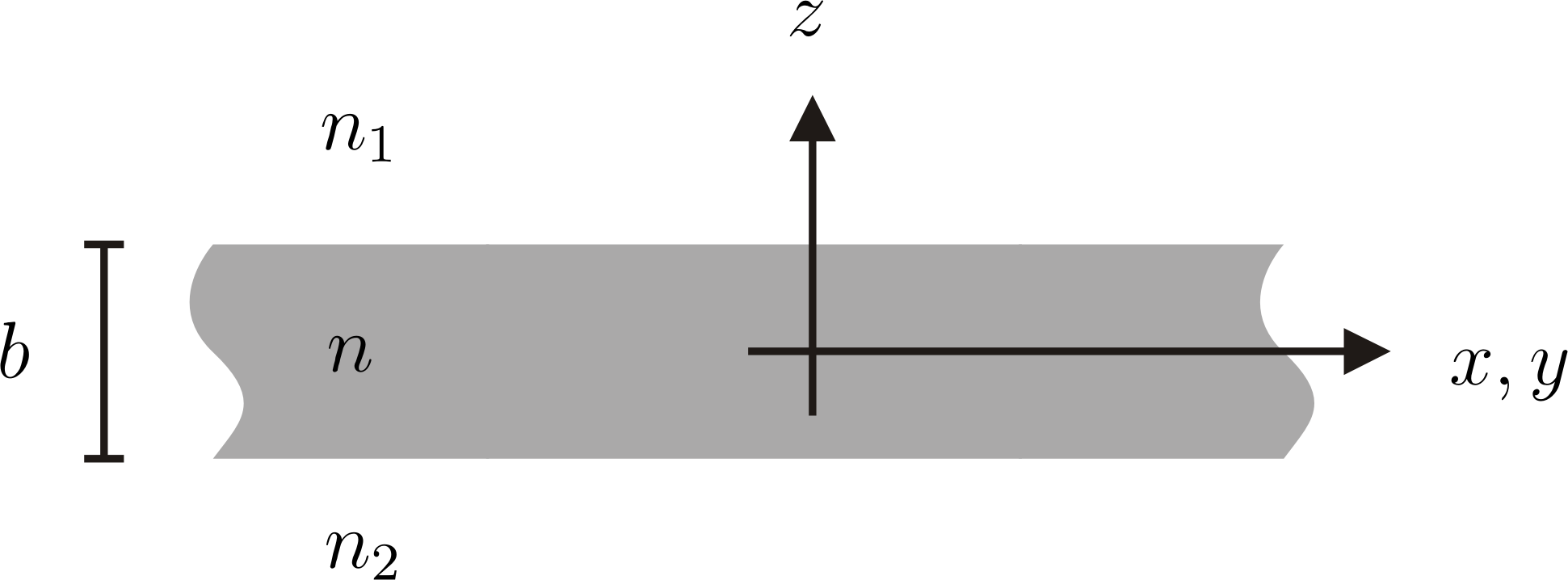}
\caption{\label{fig:infdielslabwav}Geometry and notations for the infinite dielectric slab waveguide. The dielectric slab with index of refraction $n$ and thickness $b$ is extended indefinitely in the $x$-$y$-plane and surrounded by media with index of refraction $n_1$ and $n_2$, respectively.}
\end{figure}

\begin{figure}
\includegraphics[width = 8.6 cm]{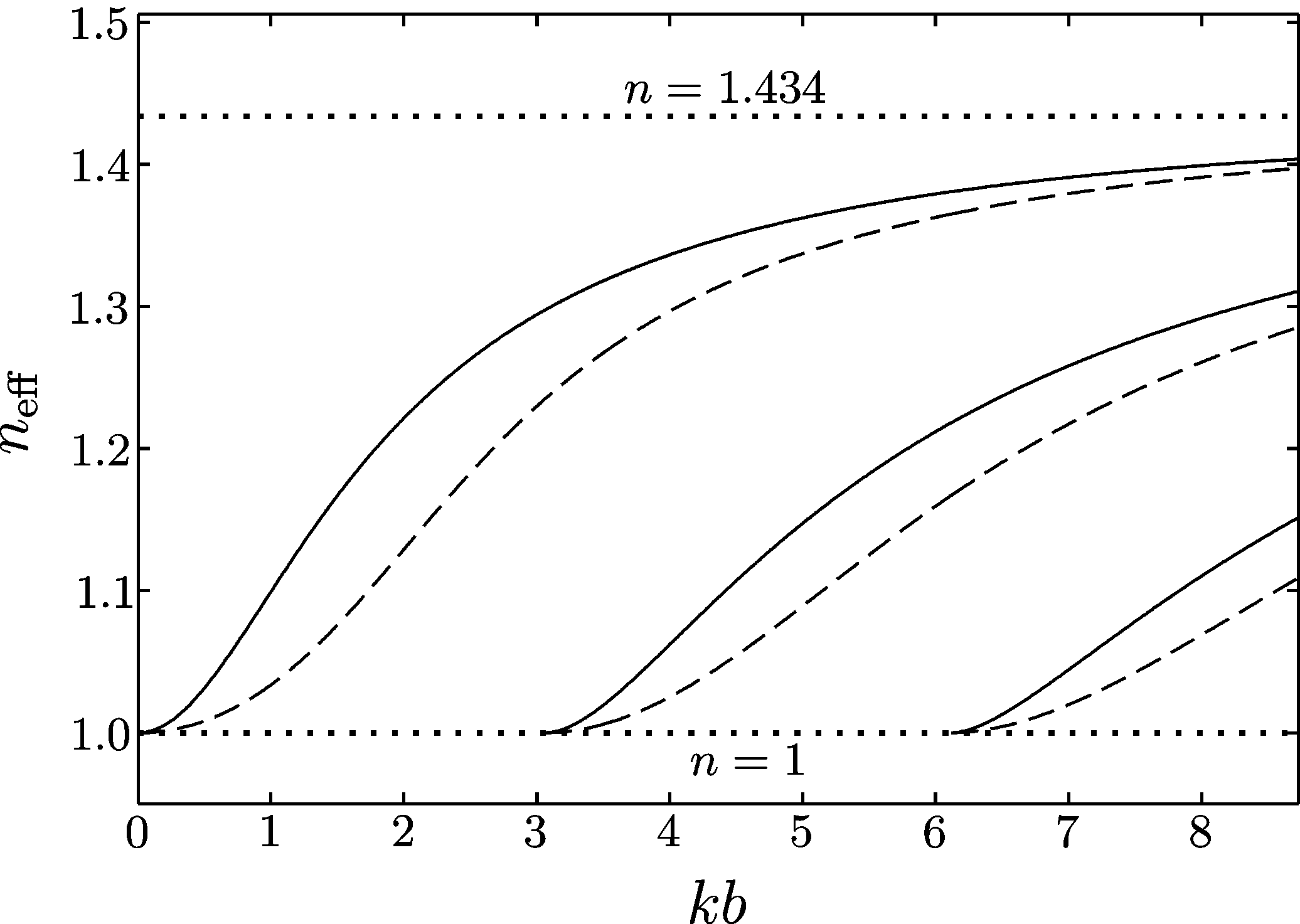}
\caption{\label{fig:neff}Effective index of refraction $\neff$ with respect to $kb$ for $n = 1.434$. The solid lines are the TE-modes, the dashed lines the TM-modes of various $z$-excitations. The dotted lines are the indices of refraction of Teflon and air, respectively.}
\end{figure}

The basic idea for the 2D-approximation is to treat the bulk of the resonator as a dielectric slab waveguide \cite{Vassallo1991}. In this section we will consider an infinite dielectric slab waveguide, that is an infinite dielectric plate of thickness $b$ with index of refraction $n$ surrounded by media with lower indices of refraction $n_{1, 2}$ (see \reffig{fig:infdielslabwav}), and treat the cylindrical sidewalls of the resonator in the next section. The surrounding media are assumed to be air ($n_{1, 2} = 1$) in the following. The waveguide is described by the vectorial Helmholtz equation
\begin{equation} \label{eq:helmholtz_vec} \left( \laplace + n^2(\bm{r}) k^2 \right) \left\{ \begin{array}{c} \bm{E} \\ \bm{B} \end{array} \right\} = \bm{0} \end{equation} where $k = \omega / c$ is the vacuum-wavenumber, $\omega$ the angular frequency and $c$ the speed of light. The Helmholtz equation (\ref{eq:helmholtz_vec}) can be simplified by separation of the variable parallel to the cylinder axis, $z$, from the $x$ and $y$ variables. Different modes in the slab waveguide can be distinguished by their polarization and their excitation perpendicular to the plane of the slab. Due to the slab geometry all field modes are either related to $E_z$ (TM-polarization) or to $B_z$ (TE-polarization) \cite{Jackson1999}. The general ansatz for $E_z$ respectively $B_z$ is 
\begin{equation} \label{eq:fields_z} \Psi(x, y) e^{-i \omega t} \left\{ \begin{array}{lcl} a_1 e^{i k_z z} + a_2 e^{-i k_z z} & : & |z| \leq b/2 \\ a_3 \exp{(-\lzt |z|)} & : & |z| \geq b/2 \end{array} \right. , \end{equation}
and the constants $a_i$ are determined from the boundary conditions. 	The wave function $\Psi$ satisfies the scalar, two-dimensional Helmholtz equation
\begin{equation} \label{eq:helmholtz_scalar} ( \laplace + \kt^2 ) \Psi = 0 , \end{equation}
with $\kt$ being the horizontal component of the wavevector. The vacuum-wavenumbers $k$ and its vertical components $k_z$ inside the dielectric medium and $\lzt$ outside are related by the dispersion relation
\begin{equation} \label{eq:disprel} \frac{\omega^2}{c^2} = k^2 = \frac{\kt^2 + k_z^2}{n^2} = \kt^2 - \lzt^2 . \end{equation}
It should be noted that $\lzt$ is real for all solutions which correspond to modes confined inside the dielectric slab by total internal reflection with evanescent fields outside. The continuity condition for $n^2(\bm{r}) E_z$ and $\pbyp{E_z}{z}$ at the interfaces at $z = \pm b/2$ for TM-modes and $B_z$ and $\pbyp{B_z}{z}$ for TE-modes yields the condition \cite{Chin1994}
\begin{equation} \label{eq:qbed_kz} k_z \tan{(k_z b/2)} = \left\{ \begin{array}{ccl} n^2 \lzt & : & \mathrm{for \, TM} \\ \lzt & : & \mathrm{for \, TE} \end{array} \right. . \end{equation}
Expressing the horizontal component $\kt$ of the wavevector in terms of the effective index of refraction $\neff$, defined as
\begin{equation} \neff = \kt / k, \end{equation}
in \refeq{eq:disprel} and \refeq{eq:qbed_kz} leads to the dispersion relation
\begin{equation} \begin{array}{c}
 k b / 2 = \frac{1}{\sqrt{n^2 - \neff^2}} \left\{ \arctan{\left( \nu \sqrt{\frac{\neff^2 - 1}{n^2 - \neff^2}} \right)} + \zeta \pi / 2 \right\} \\ \mathrm{with} \quad \nu = \left\{ \begin{array}{ccl} n^2 & : & \mathrm{for \, TM}_\zeta \\ 1 & : & \mathrm{for \, TE}_\zeta \end{array} \right. 
\end{array} \end{equation}
for $\neff$ with $\zeta = 0, 1, 2, \dots$ denoting the order of excitation in $z$-direction. Since in the following only $\mathrm{TM}_0$ and $\mathrm{TE}_0$ modes are considered, we skip the index. Obviously, $\neff$ depends only on the index of refraction $n$ and on $k b \propto b / \lambda$, that means on the ratio of the slab's thickness $b$ to the wavelength $\lambda$. A plot of $\neff$ with respect to $kb$ is shown in \reffig{fig:neff}. The effective index of refraction is always $1 \leq \neff \leq n$. Modes with $z$-excitation $\zeta$ emerge at the cutoff $kb = \zeta \pi / \sqrt{n^2 - 1}$.

\section{\label{sec:dielcircle}Application to Dielectric Resonators}
The solution of $\neff$ is now inserted into \refeq{eq:helmholtz_scalar}, leading to the two-dimensional scalar Helmholtz equation
\begin{equation} ( \laplace + \neff^2 k^2 ) \Psi = 0 \end{equation}
with $\Psi$ corresponding to $E_z$ ($B_z$) for TM- (TE-) modes. This equation correctly describes the propagation of electromagnetic waves inside an infinite dielectric slab. The next step is to incorporate the cylindrical sidewalls of the dielectric resonator in a plane perpendicular to the cylinder axis. In the 2D approximation this is achieved by considering the vertical walls as a part of an infinite dielectric cylinder with $n = \neff$ and imposing the corresponding boundary conditions \cite{Jackson1999}, i.e.\ that fields inside and outside the resonator obey the Helmholtz equation
\begin{equation} \label{eq:helmholtz_neff} \laplace \Psi_{\mathrm{in, out}} = \left\{ \begin{array}{lcl} - \neff^2 k^2 \Psi_{\mathrm{in}} & : & \bm{r} \in S \\ - k^2 \Psi_{\mathrm{out}} & : & \bm{r} \notin S . \end{array} \right. \end{equation}
Here $S$ is the domain of the resonator in a plane perpendicular to the cylinder axis, and the conditions along the boundary $\partial S$ of $S$
\begin{equation} \label{eq:neff_bcs} \Psi_{\mathrm{in}}|_{\partial S} = \Psi_{\mathrm{out}}|_{\partial S} \quad \mathrm{and} \quad \mu \left. \pbyp{\Psi_{\mathrm{in}}}{\bm{n}} \right|_{\partial S} = \left. \pbyp{\Psi_{\mathrm{out}}}{\bm{n}} \right|_{\partial S} \end{equation}
are imposed, with $\mu$ equal to $1$ ($1 / \neff^2$) for TM- (TE-) modes and $\bm{n}$ being the unit vector normal to the surface. The set of equations (\ref{eq:helmholtz_neff}) and (\ref{eq:neff_bcs}) constitutes the quintessence of the $\neff$-model. It correctly describes a two-dimensional dielectric resonator, i.e. the electromagnetic field is homogeneous in the $z$-direction, with index of refraction equal to $\neff$, however, not a flat but three-dimensional resonator. Indeed the full 3D Maxwell equations lead to additional boundary conditions if the electromagnetic fields depend on $z$ \cite{Schwefel2005, Dubertrand2008}, which couple the TM- and TE-polarizations. Analytical calculations incorporating these have to our knowledge not yet been performed. As has already been stated above, the aim of the present work is to test experimentally the applicability of the $\neff$-model and to understand the order of magnitude of deviations from the experiment. \newline
The solutions of Eqs. (\ref{eq:helmholtz_neff}) and (\ref{eq:neff_bcs}) for a circular resonator with radius $R$ are given in cylindrical coordinates $r$ and $\varphi$ as
\begin{equation} \Psi_{\mathrm{in}}(r, \varphi) = \Psi_{\mathrm{in}}^{(0)} \besselj{m}{\neff k r} e^{\pm i m \varphi} \end{equation}
inside the disk and as
\begin{equation} \Psi_{\mathrm{out}}(r, \varphi) = \Psi_{\mathrm{out}}^{(0)} \hankel{m}{1}{k r} e^{\pm i m \varphi} \end{equation}
outside. Here, $\besselj{m}{x}$ is a Bessel-function of the first kind, $\hankel{m}{1}{x}$ a Hankel-function of the first kind and $m$ the azimuthal quantum number. All modes with $m > 0$ are doubly degenerate. Applying the boundary conditions (\refeq{eq:neff_bcs}) leads to the quantization condition \cite{Hentschel2002b}
\begin{equation} \label{eq:qbed_diel_circ} \mu \, \neff \frac{\besseljp{m}{\neff k R}}{\besselj{m}{\neff k R}} = \frac{\hankelp{m}{1}{kR}}{\hankel{m}{1}{kR}} , \end{equation}
where $\besseljp{m}{x}$ and $\hankelp{m}{1}{x}$ are the derivatives with respect to $x$. For each azimuthal quantum number $m$, there is an infinite series of complex solutions $k_{m, n_r}$; $n_r = 1, 2, \dots$ is the radial quantum number. Since a dielectric resonator described by equations (\ref{eq:helmholtz_neff}) and (\ref{eq:neff_bcs}) is an open system, its modes have losses due to radiation. These are determined by the imaginary part of $k_{m, n_r}$, and the quality factor $Q$ of a mode is defined as
\begin{equation} Q = - \frac{\Re{k_{m, n_r}}}{2 \, \Im{k_{m, n_r}}} . \end{equation}
Because of these losses, the solutions of \refeq{eq:helmholtz_neff} are called quasi-bound modes. They are identified as the poles of the scattering matrix $S$ describing the measurement process \cite{Noeckel2002} --- microwave power is coupled into the resonator via an antenna, thereby exciting modes, and coupled out via the same or another antenna --- in the form of resonances in the frequency spectrum. \newline
The quantization condition \refeq{eq:qbed_diel_circ} is solved numerically, taking into account the dependence of $\neff$ on (the real part of) $k$.

\section{\label{sec:expsetup}Experimental Technique}

\begin{figure}[tbh]
\includegraphics[width = 8.6cm]{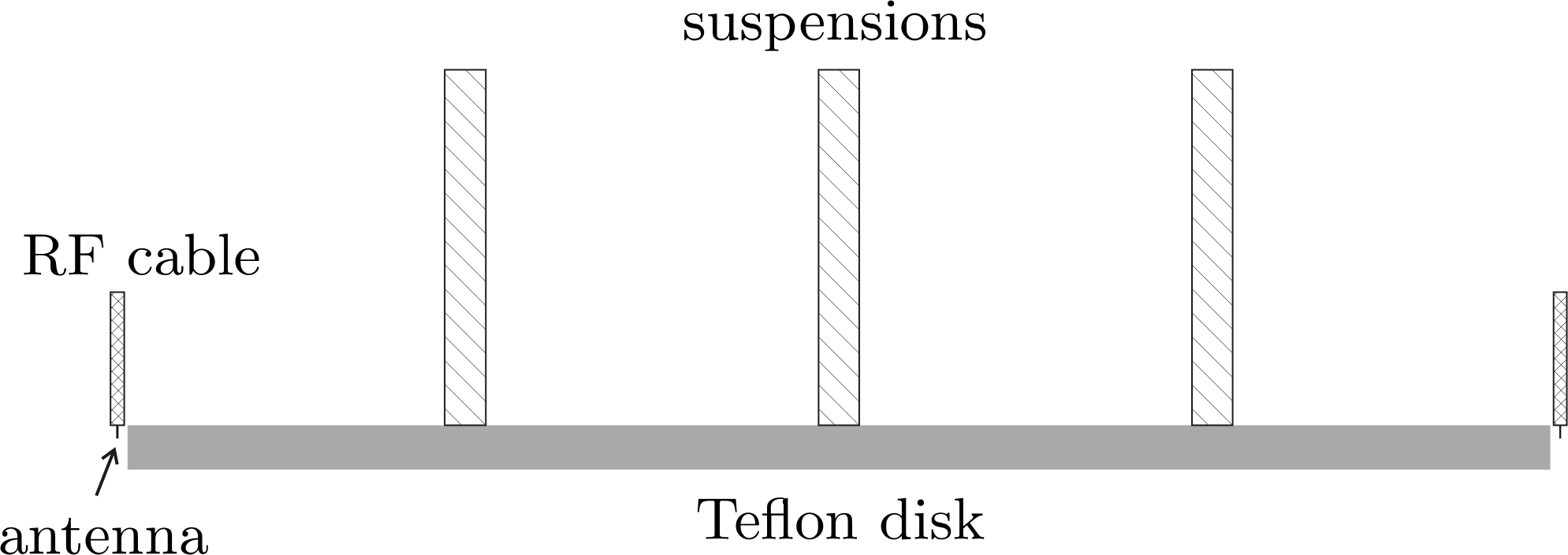}
\caption{\label{fig:setup}Schematic side view of the experimental setup. The Teflon disk is hanging on three metal suspensions. Two dipole antennas protruding from an RF cable are placed close to the rim of the disk on opposite sides.}
\end{figure}

\begin{figure}[tbh]
\includegraphics[width = 8.6 cm]{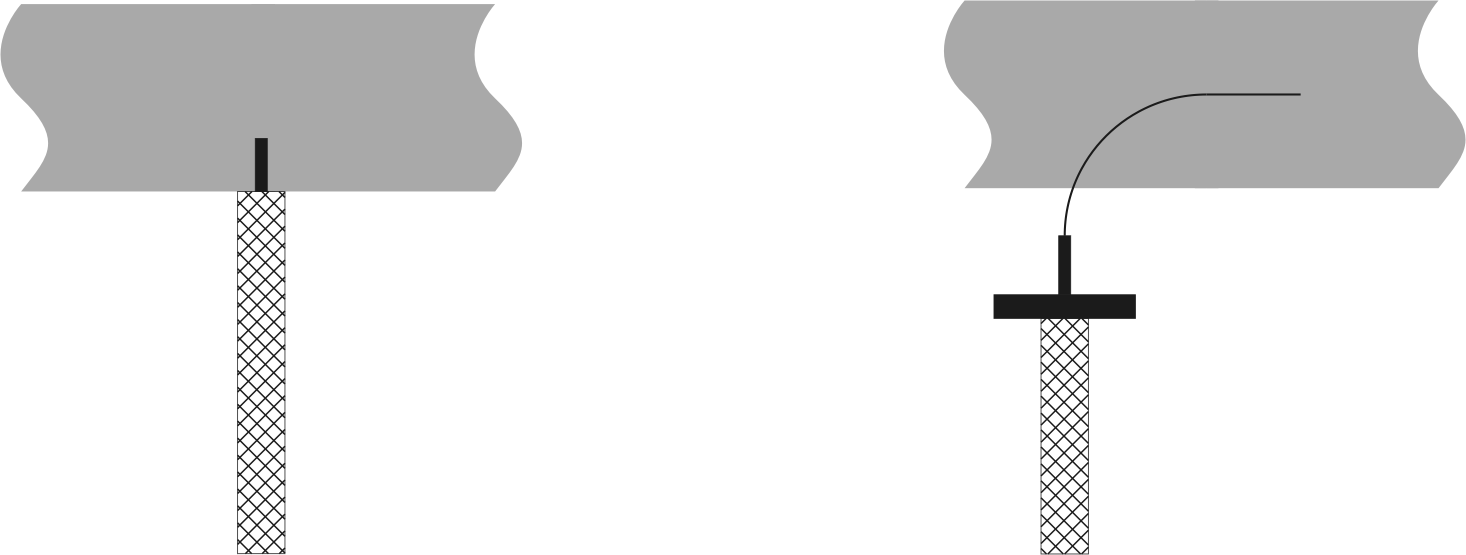}
\caption{\label{fig:ants}Sketches of the two antenna types. A dipole antenna is shown on the left, a curved antenna on the right. The cylindrical sidewall of the Teflon disk is shown in gray in the background. Both types of antennas are placed directly alongside the sidewall of the disk to obtain good coupling to the resonator.}
\end{figure}

\begin{figure*}[tbh]
\includegraphics[width = 16 cm]{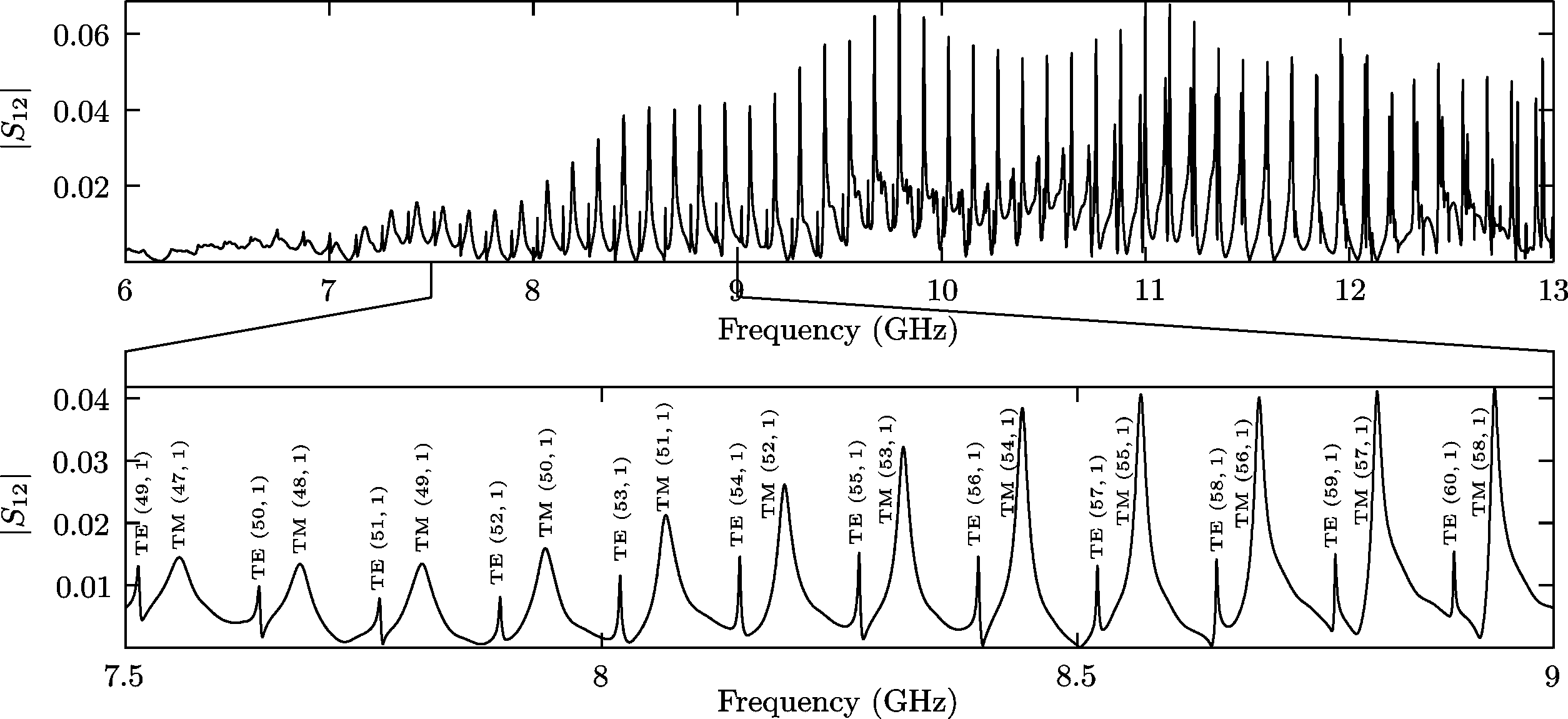}
\caption{\label{fig:spec_HKreis}Frequency spectrum of disk A measured with dipole antennas. The transmission amplitude, i.e.\ the modulus $|S_{12}|$, is shown with respect to the frequency. In the magnified part of the spectrum, the resonances are labeled with TM/TE $(m, n_r)$ to indicate their polarization as well as their azimuthal and radial quantum numbers $m$ and $n_r$, respectively. Two series of resonances can be seen here: The broader and larger resonances correspond to modes with TM-polarization and radial quantum number $n_r = 1$, the sharper and smaller resonances to modes with TE-polarization and $n_r = 1$. Resonances with $n_r > 1$ can also be seen at higher frequencies.}
\end{figure*}

\begin{figure*}[tbh]
\begin{center}
\begin{minipage}{5.4 cm}
	\begin{center}
		\includegraphics[width = 5.3 cm]{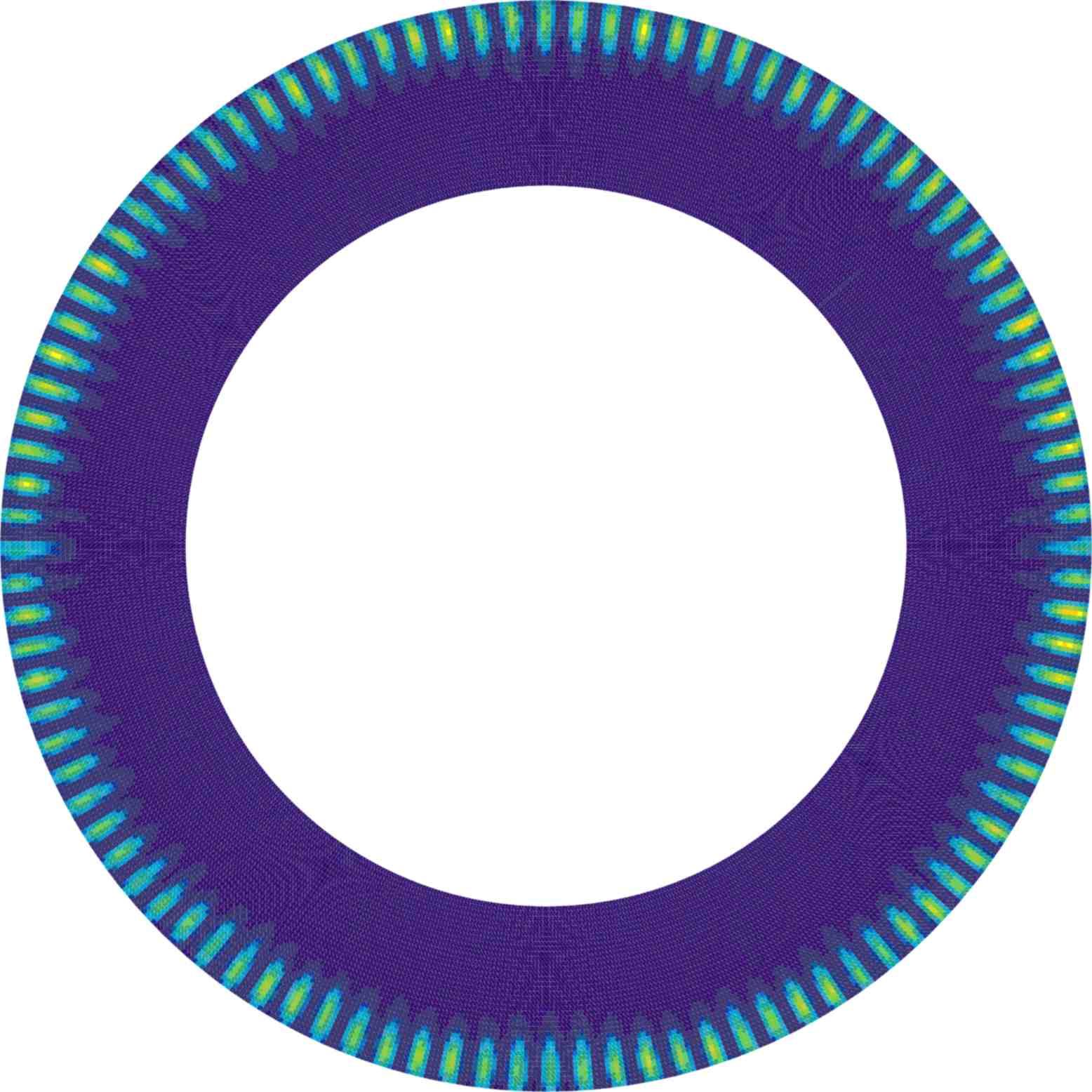} 
		$7.94$ GHz: TM $(50, 1)$
	\end{center}
\end{minipage}
\begin{minipage}{5.4 cm}
	\begin{center}
		\includegraphics[width = 5.3 cm]{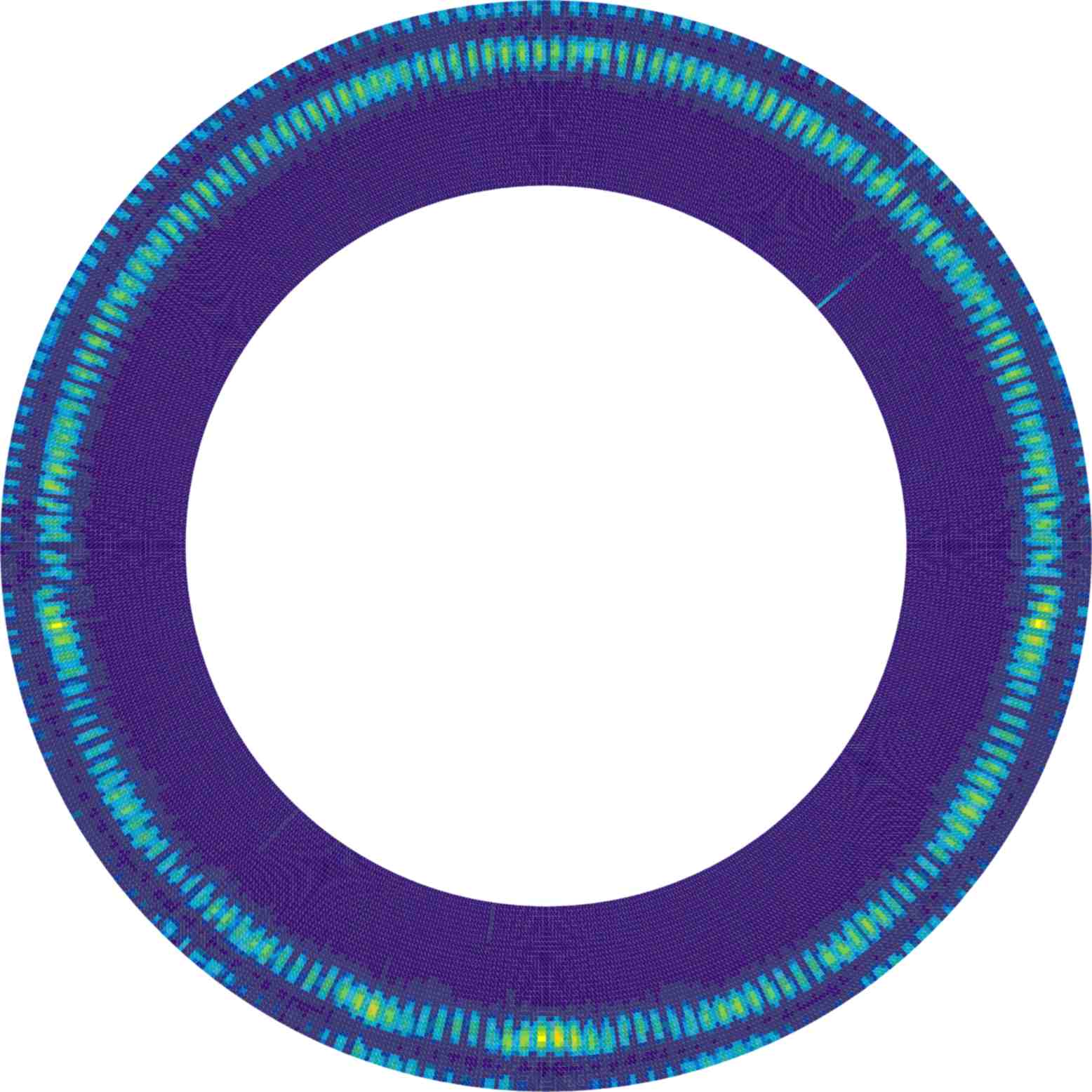}
		$13.80$ GHz: TM $(92, 2)$
	\end{center}
\end{minipage}
\begin{minipage}{5.4 cm}
	\begin{center}
		\includegraphics[width = 5.3 cm]{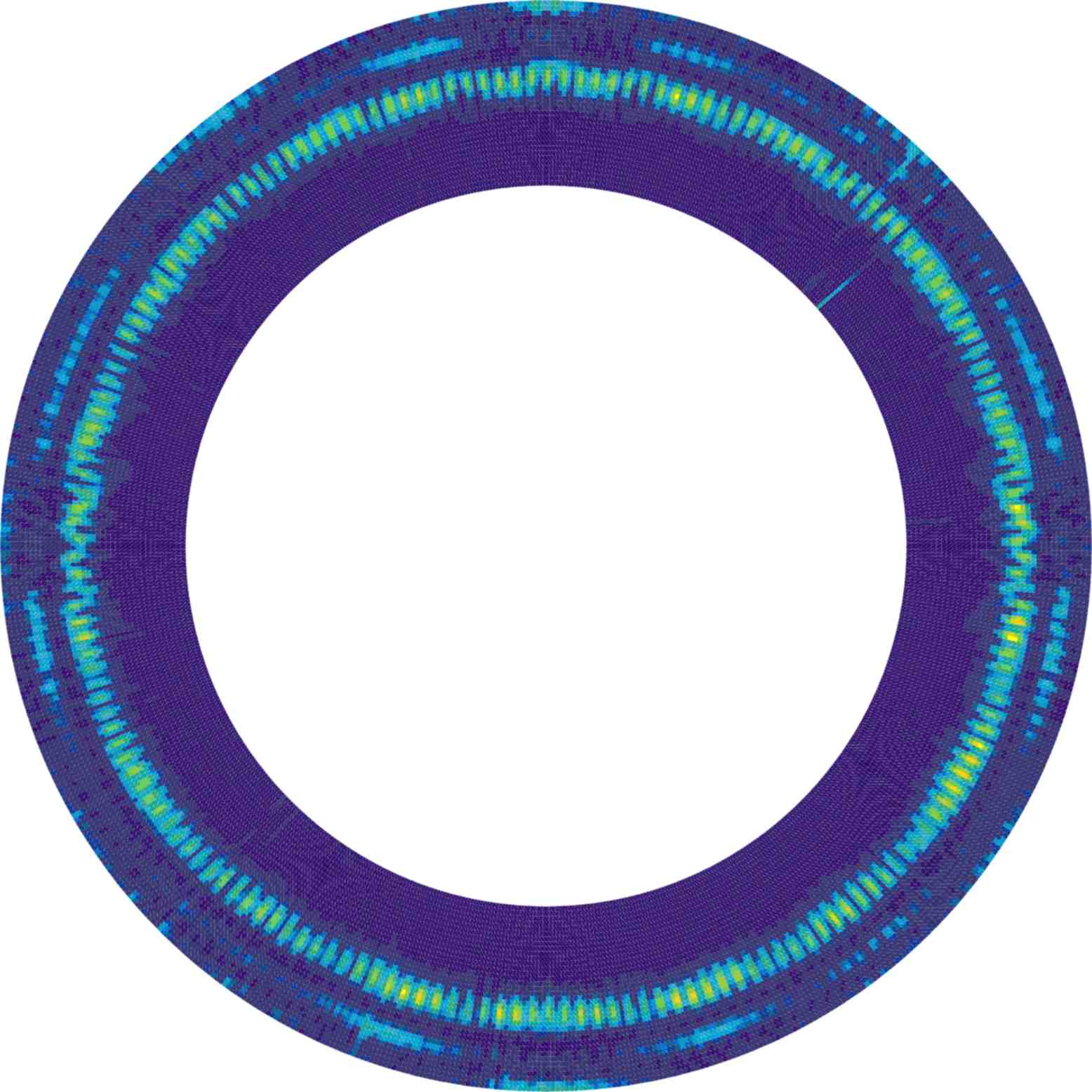}
		$14.33$ GHz: TM $(91, 3)$
	\end{center}
\end{minipage}
\begin{minipage}{1 cm}
	\begin{center}
		\includegraphics[height = 5.4 cm]{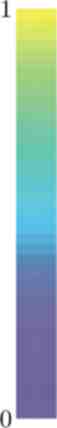}
		\vspace{0.5 cm}
	\end{center}
\end{minipage}
\end{center}
\caption{\label{fig:WFs}(Color online) Measured intensity distributions of three TM-modes with quantum numbers $(m, n_r)$. A mode with azimuthal quantum number $m$ and radial quantum number $n_r$ has $2 m$ maxima in azimuthal direction and $n_r$ rings. Shown are modes of whispering gallery type, as are all other identifiable modes. Therefore, intensities were measured only in the outer ring of the resonator.}
\end{figure*}

The experimental setup is sketched in \reffig{fig:setup}: The Teflon disk is hanging on three metal suspensions attached to the disk at the corners of an equilateral triangle, but otherwise surrounded only by air. Two antennas are put on opposite sides of the disk. The whole setup is placed in a thermostat to keep it at a fixed temperature. It should be noted that the perturbation of the resonator due to the attached suspensions leads to negligible changes of the resonance frequencies by less than 2\,\% of the mean resonance spacing. The perturbations due to the bending of the resonator under its own weight and the antennas are of the same order of magnitude. Two different types of antennas (see \reffig{fig:ants}) were used: vertical dipole antennas, which excite mainly TM-modes, and so-called curved antennas, which excite mainly TE-modes. The antennas may slightly lift the degeneracy of the modes. A vectorial network analyzer (PNA N5230A by Agilent Technologies) is used to measure the scattering matrix element $S_{21}(f)$, where the modulus squared of $S_{21}(f)$ is the ratio
\begin{equation} |S_{21}(f)|^2 = \frac{P_{\mathrm{out}}}{P_{\mathrm{in}}} \end{equation}
between the power $P_{\mathrm{in}}$ coupled in by antenna $1$ and the power $P_{\mathrm{out}}$ coupled out by antenna 2 for a given frequency $f = \omega / 2 \pi$. \newline
Two circular disks of different thickness $b$ made of Teflon were used in the experiments to investigate the dependence of the resonance frequencies on the thickness. Disk A has a radius of \mbox{$R = 274.8$ mm} and a thickness of \mbox{$b = 16.7$ mm}, and disk B of \mbox{$R = 274.9$ mm} and of \mbox{$b = 5.0$ mm}. A frequency of $10$ GHz corresponds to $kR \approx 57.6$, and $kb \approx 3.5$ (disk A) and $kb \approx 1.0$ (disk B), respectively. The index of refraction was measured using a split-cylinder resonator \cite{Janezic2004a, Kent1988} and is \mbox{$n = 1.434 \pm 0.01$} for disk A and \mbox{$n = 1.439 \pm 0.01$} for disk B. An example of a measured frequency spectrum of disk A (with dipole antennas) is shown in \reffig{fig:spec_HKreis}. It displays a superposition of several series of almost equidistant resonances, each corresponding to modes with a fixed polarization and radial quantum number $n_r$ and ascending azimuthal quantum number $m$. The resonance spacing for each of these subspectra is typically \mbox{$120$--$130$ MHz}. This is illustrated in the lower part of \reffig{fig:spec_HKreis}. Within each subspectrum, the width of the resonances decreases with increasing azimuthal quantum number. This can be explained within the ray picture \cite{Hentschel2002b}: A higher azimuthal quantum number corresponds to a higher angular momentum, and this to rays with a larger angle of incidence at the boundary, implying lower radiation losses. As a consequence, different subspectra are distinguishable only above a certain frequency and contribute below only to the background. The dependence of the amplitude of a resonance on the type of antenna used gives a hint at the polarization of the mode. It was determined in addition with a perturbation technique: A metal plate was introduced parallel to the Teflon disk (with varying distance) leading to a shift of the resonance frequencies. Due to the different boundary conditions for TE- (Dirichlet for $B_z$) and TM- (Neumann for $E_z$) modes at the metal plate the former are shifted to lower, the latter to higher frequencies with decreasing distance between the metal plate and the disk. The quantum numbers were determined from the intensity distributions, which were measured with the perturbation body method \cite{Maier1952}. A cylinder made of magnetic rubber \cite{Bogomol2006} was used as perturbation body and moved along the surface of the disk, its height of \mbox{$8$ mm} and diameter of \mbox{$4$ mm}  being small compared to a vacuum-wavelength of \mbox{$30$ mm} at \mbox{$10$ GHz}. Then the positioning of the perturbation body on the disk leads to a shift of the resonance frequency which is proportional to the electric field intensity at its position. Three examples are shown in \reffig{fig:WFs}. All the measured modes are of whispering gallery mode type. As a result the resonances are only slightly perturbed by the suspensions, since they are located well inside the caustic of the whispering gallery modes. Accordingly, the intensities were measured only in the outer part of the disk.  With the knowledge of the polarization and the quantum numbers, the measured resonance frequencies can be compared with those computed based on the $\neff$-model (solutions of \refeq{eq:qbed_diel_circ}).

\section{\label{sec:HKreis}Comparison of model and experiment for disk A}

\begin{figure}[bth]
\subfigure[\,TE-modes]{
	\includegraphics[width = 8.4 cm]{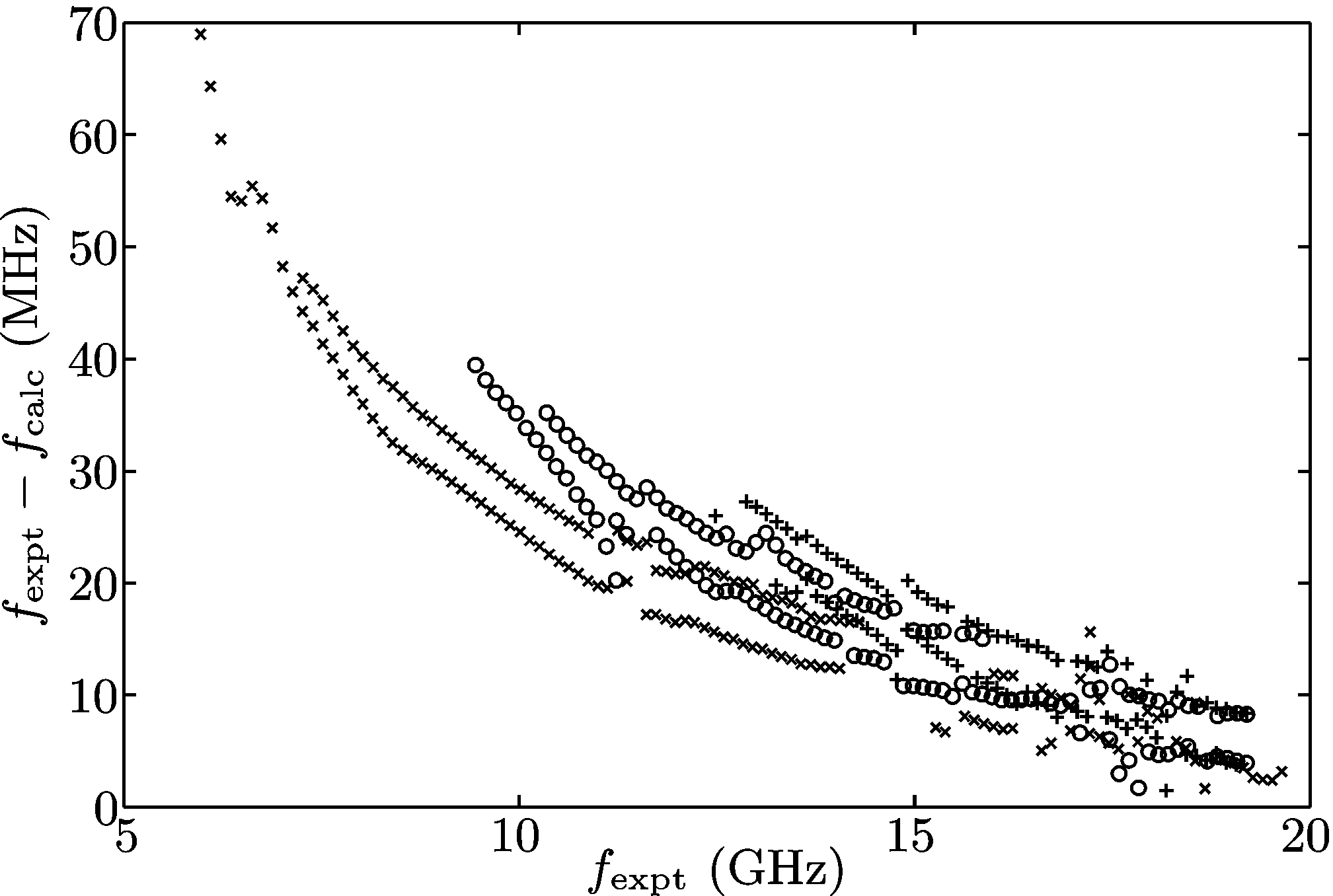}
	\label{subfig:fe-ft_HKreis_TE}
}
\subfigure[\,TM-modes]{
	\includegraphics[width = 8.4 cm]{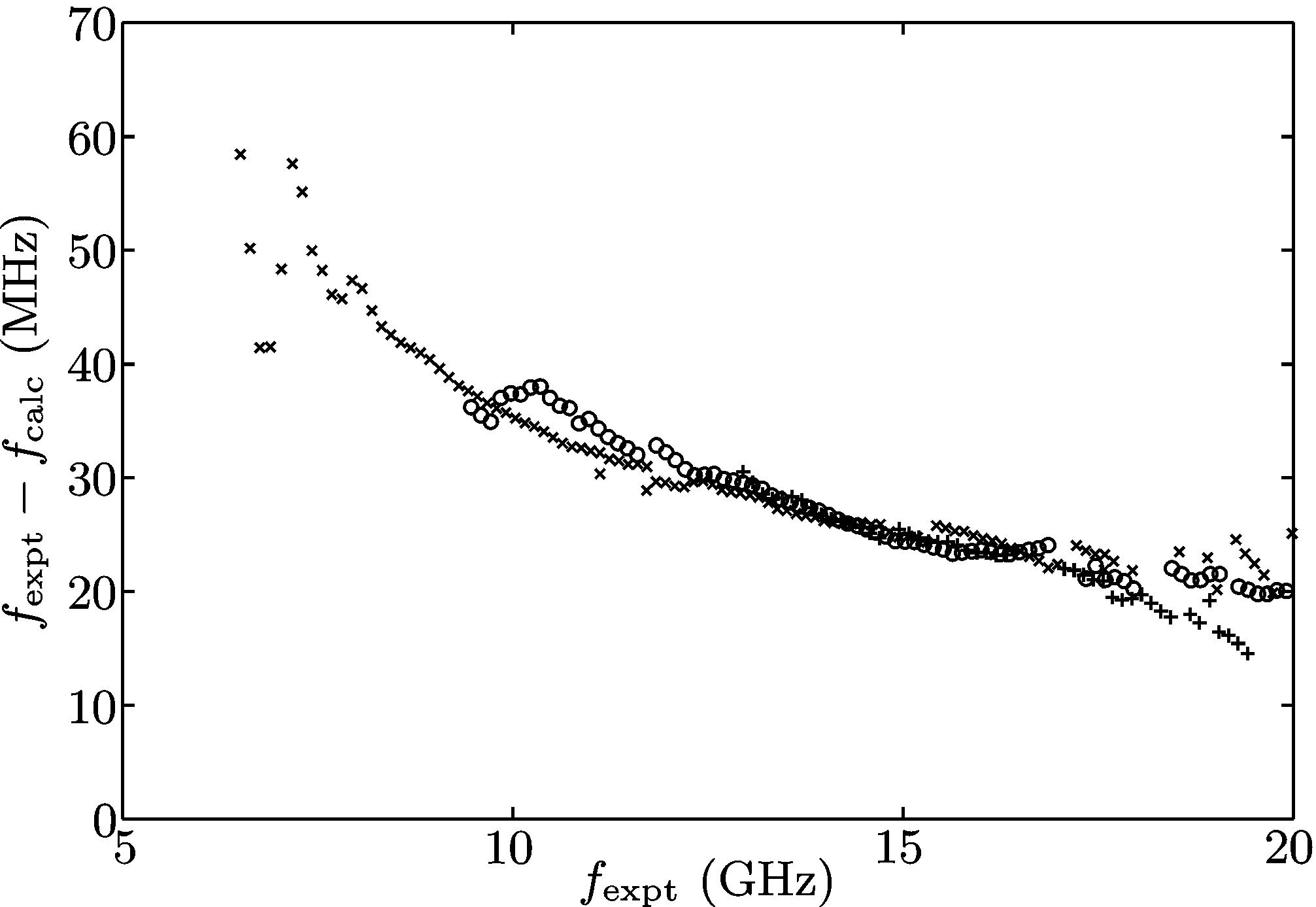}
	\label{subfig:fe-ft_HKreis_TM}
}
\caption{\label{fig:fe-ft_HKreis}Difference between measured ($\fexp$) and calculated ($\ftheo$) resonance frequencies with respect to $\fexp$ for disk A. The different symbols correspond to the different radial quantum numbers ($\times$: $n_r = 1$, $\circ$: $n_r = 2$, $+$: $n_r = 3$). \subref{subfig:fe-ft_HKreis_TE} TE-modes: The range of azimuthal quantum numbers for $n_r = 1$ is $m = 37$--$148$. For each $n_r$, there are two series of data points due to the break-up of the degenerate modes by the curved antenna used in the measurement. The frequencies of the unperturbed system are approximately in between. \subref{subfig:fe-ft_HKreis_TM} TM-modes: The range of azimuthal quantum numbers for $n_r = 1$ is $m = 39$--$150$. The measurement was done with dipole antennas, and no break-up of degenerate modes was observed.}
\end{figure}

\begin{figure}[tbh]
\includegraphics[width = 8.6 cm]{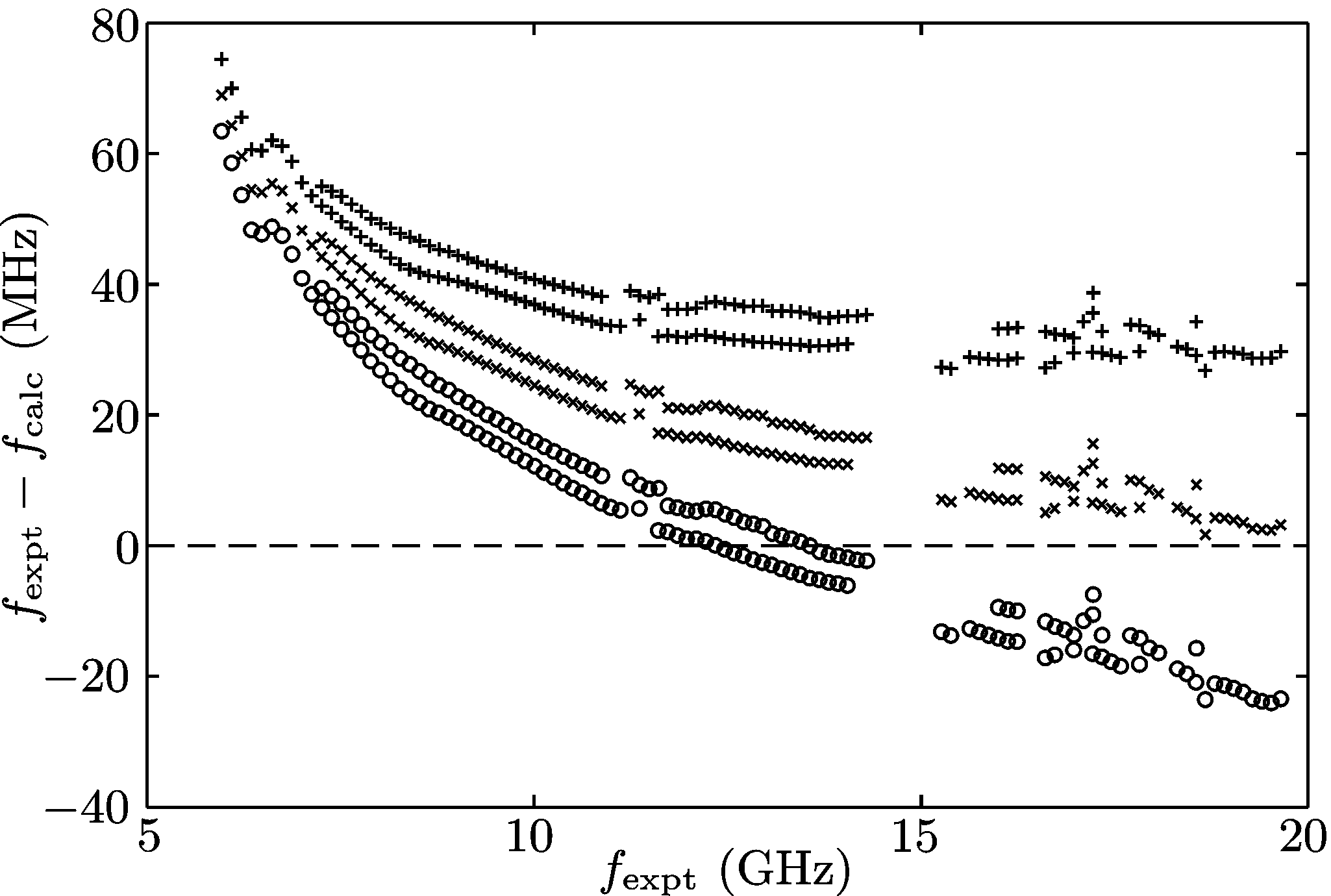}
\caption{\label{fig:fe-ft_3n}Difference between measured ($\fexp$) and calculated ($\ftheo$) resonance frequencies with respect to $\fexp$ for the TE-modes of disk A. Results from three different calculations with three different indices of refraction $n = 1.432 \, (\circ), \, 1.434 \, (\times)$ and $1.436 \, (+)$ are shown only for resonances with $n_r = 1$ for the sake of clarity. The difference $\fexp - \ftheo$ depends strongly on $n$ in the semiclassic limit, i.e.\ for high frequencies, but only very slightly for low frequencies.}
\end{figure}

\begin{figure}[tbh]
\subfigure[\,TE-modes]{
	\includegraphics[width = 8.4 cm]{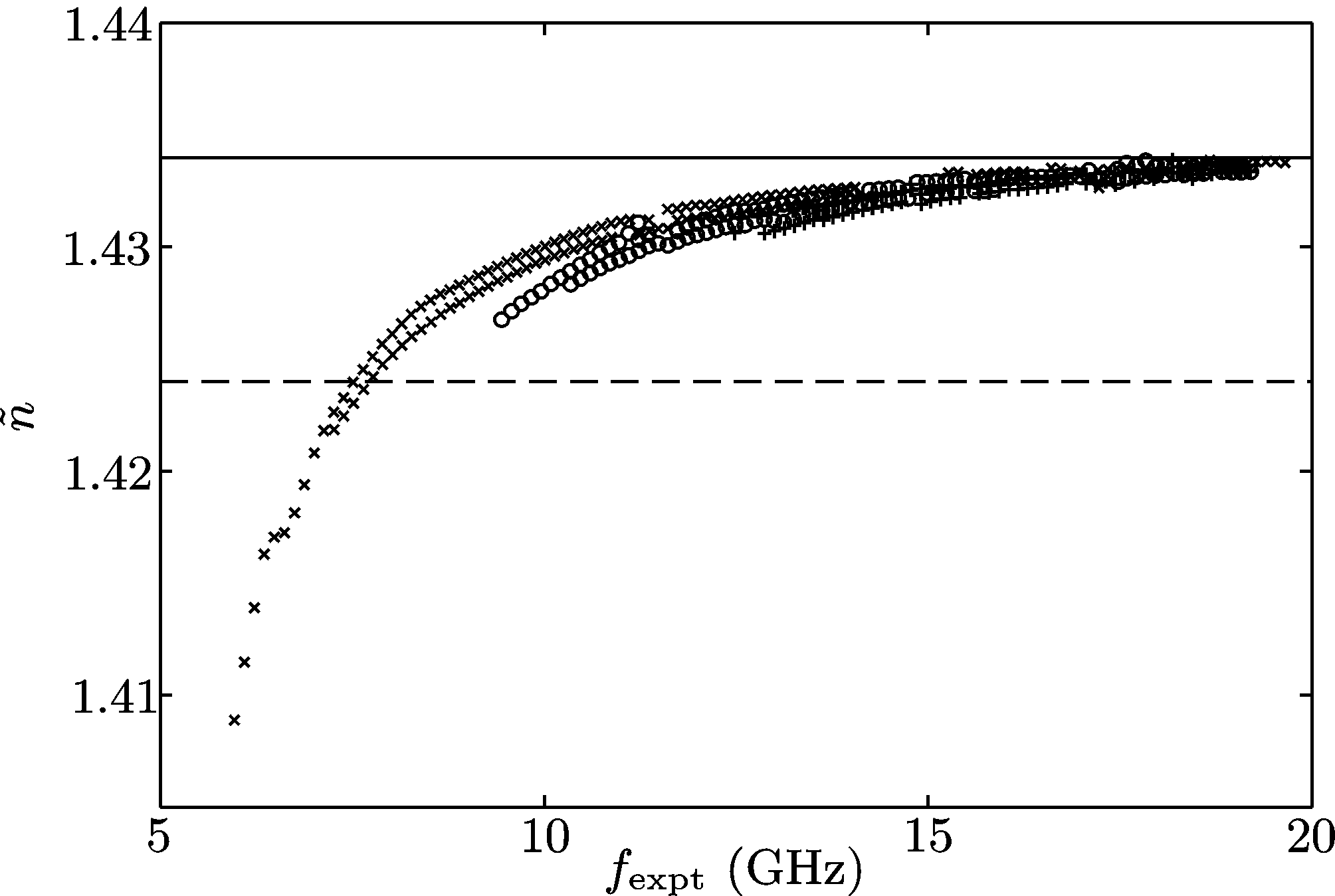}
	\label{subfig:nexp_HKreis_TE}
}
\subfigure[\,TM-modes]{
	\includegraphics[width = 8.4 cm]{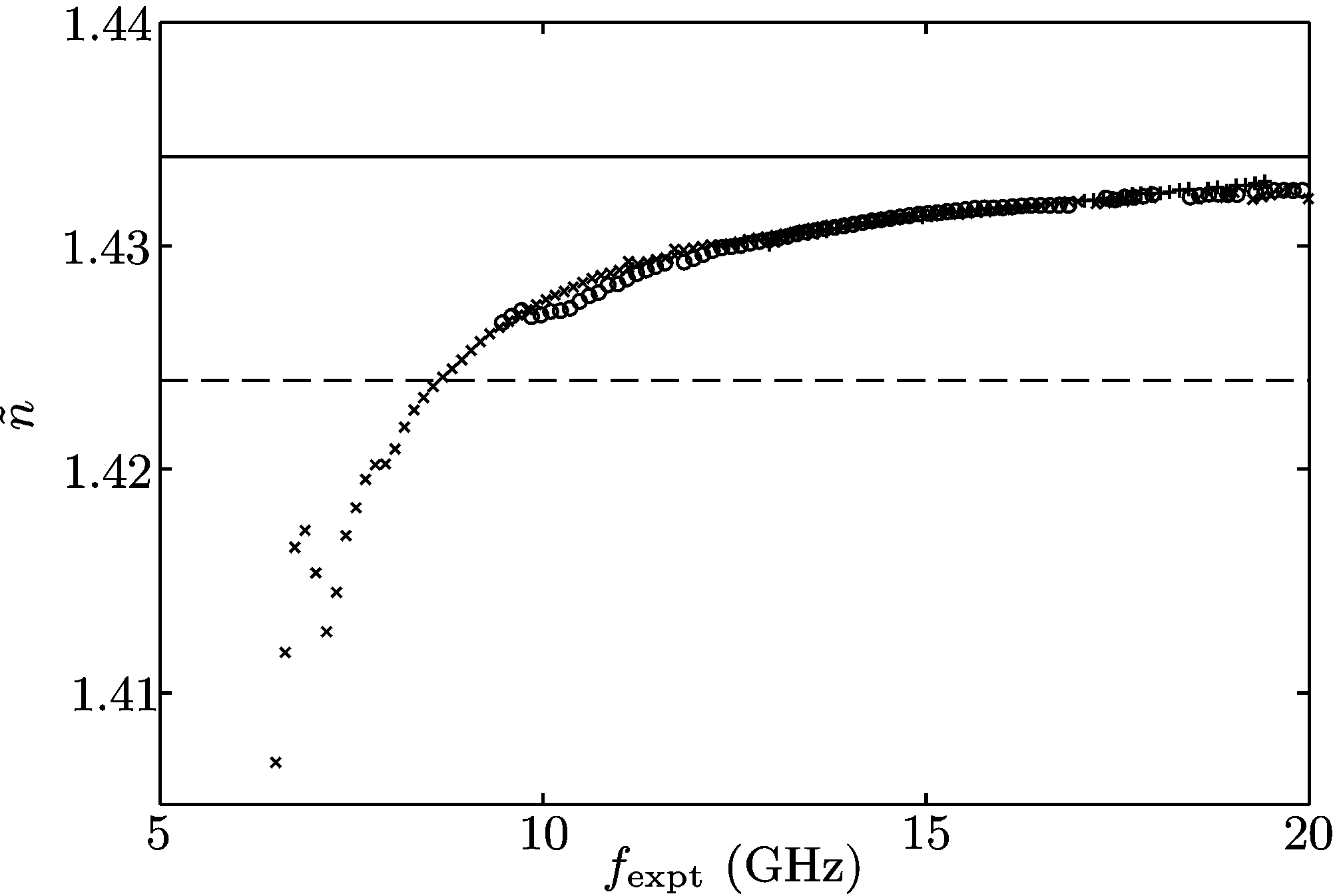}
	\label{subfig:nexp_HKreis_TM}
}
\caption{\label{fig:nexp_HKreis} Values of the index of refraction $\nexp$ for which \refeq{eq:qbed_diel_circ} yields the measured resonance frequencies $\fexp$ with respect to the measured resonance frequency for disk A. The upper graph \subref{subfig:nexp_HKreis_TE} shows the result for the TE-modes, the lower one \subref{subfig:nexp_HKreis_TM} that for the TM-modes. The different symbols denote the different radial quantum numbers ($\times$: $n_r = 1$, $\circ$: $n_r = 2$, $+$: $n_r = 3$). The solid line is the real index of refraction $n$ of the disk, the dashed line $n - \Delta n$. The systematic deviation of the data points from the measured $n$ shows the failure of the $\neff$-model.}
\end{figure}

\begin{figure}[tbh]
\includegraphics[width = 8.6 cm]{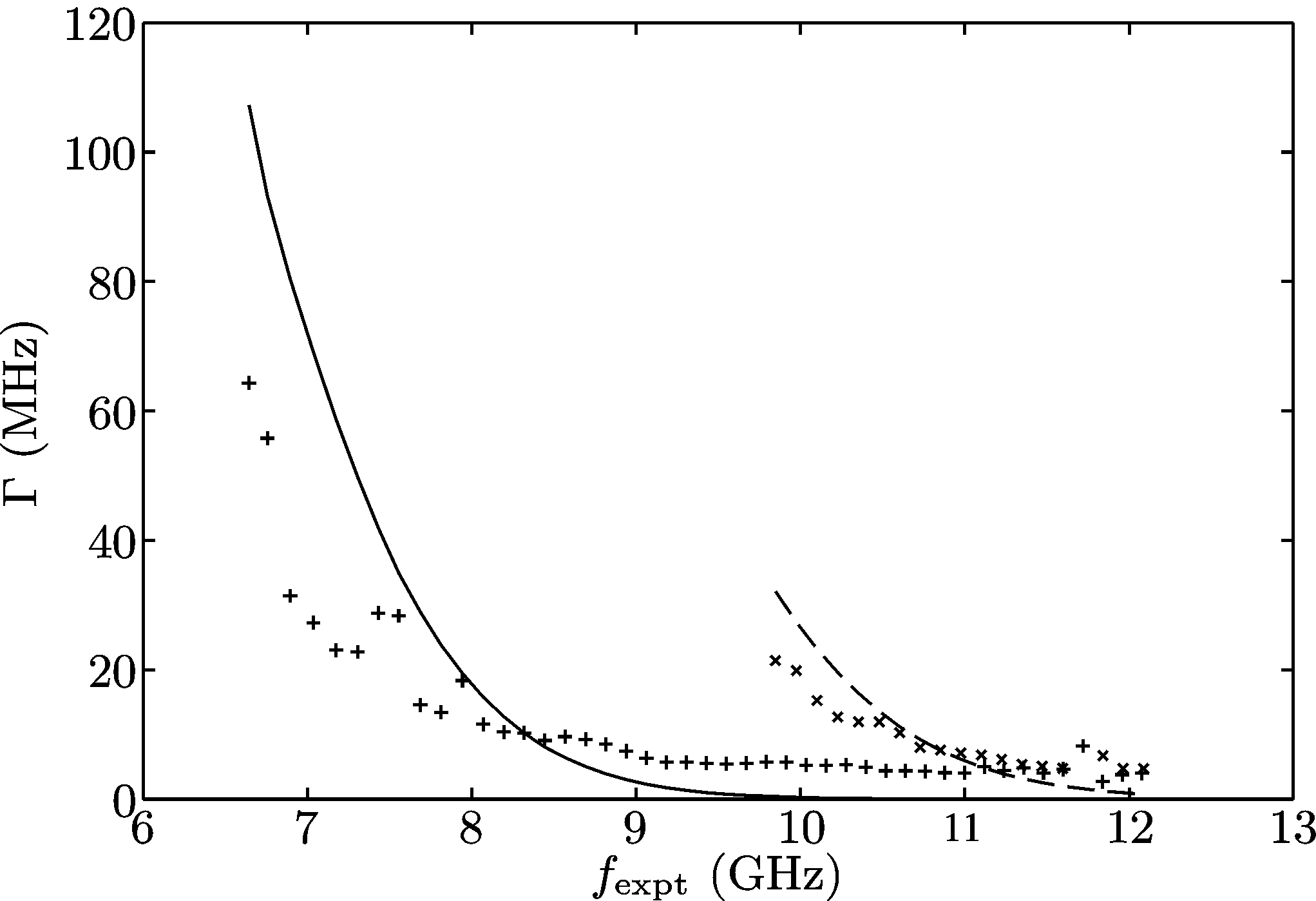}
\caption{\label{fig:Gamma_HKreis}Measured ($\Gexp$) and calculated ($\Gtheo$) resonance widths for TM-modes of disk A. Both $\Gexp$ and $\Gtheo$ are plotted as function of the measured resonance frequency $\fexp$. The different symbols denote the measured widths for different radial quantum numbers ($+$: $n_r = 1$, $\times$: $n_r = 2$). The calculated widths are plotted as curves (solid line for $n_r = 1$, dashed for $n_r = 2$) instead of data points to guide the eye.}
\end{figure}

Figure \ref{fig:fe-ft_HKreis} shows the difference between the measured resonance frequencies $\fexp$ of disk A (\mbox{$b = 16.7$ mm} thick) and those calculated with \refeq{eq:qbed_diel_circ}, $\ftheo$. In \reffig{subfig:fe-ft_HKreis_TE} (TE-modes), there are two series of data points for each radial quantum number $n_r$ because the degeneracy of the modes is lifted by the curved antennas used in the measurement. The scattering of the data points of about \mbox{$\pm 5$ MHz} in both graphs stems from problems with the determination of the resonance frequencies, either because the resonances are badly shaped (at lower frequencies) or because of overlapping resonances (at higher frequencies). Some resonances are not detectable due to the overlap with others. The deviation between the measured and computed resonance frequencies is generally less than 1\,\%. Still, these deviations of the model calculations from the experimental data must be considered significant: A deviation of $60$ MHz is about half the resonance spacing between resonances with the same $n_r$, and thus it is impossible to correctly identify the different resonances just by a comparison with the model calculations. The difference between the calculated and the measured resonance frequencies decreases with increasing frequency for both polarizations and seems to reach a finite value in both cases. Interestingly, this value is different for TE- and TM-modes. Furthermore, the magnitude of the deviations slightly depends on the radial quantum number, especially for the TE-modes. All this hints at a systematic failure of the model, even in the semiclassical, i.e.\ high frequency limit, although the decrease of the deviations with increasing frequency indicates that the $\neff$-model is more accurate in the semiclassical regime. It should be noted that the magnitude of deviations between model and experiment is extremely sensitive to the exact value of the index of refraction, which has been determined with an uncertainty \mbox{$\Delta n = \pm 0.01$}. In order to illustrate that this accuracy is insufficient for an exact determination of the deviations, $\fexp - \ftheo$ is plotted in \reffig{fig:fe-ft_3n} for calculations with three values of $n$, which differ by even less than $\Delta n$. Only data points with $n_r = 1$ are shown, the data points for resonances with $n_r > 1$ go along with them. The three calculations have a very different behavior in the semiclassical limit, but roughly agree for low frequencies. \newline
In a next step we considered the index of refraction $n$ as a fit parameter --- let's call it $\nexp$ --- in \refeq{eq:qbed_diel_circ}, where $\nexp$ enters implicitly via $\neff$, and varied it such that \refeq{eq:qbed_diel_circ} yields the measured resonance frequencies. If the $\neff$-model provides a good description, the resulting values of $\nexp$ should scatter around the actual value of $n$. The results are shown in \reffig{fig:nexp_HKreis}. Almost all data points (except the ones at low frequencies) lie inside the error band $n - \Delta n$, but they form three distinct curves for the three values of the radial quantum number $n_r$. This provides further evidence for the failure of the $\neff$-model: If the model were correct, the values of $\nexp$ would form a single line along the actual index of refraction $n$ of the disk for both polarizations and all three values of $n_r$. It is known from literature (and was confirmed experimentally) that Teflon has negligible dispersion in this frequency range. In conclusion the observed deviations between the model and the experiment cannot be attributed to a badly determined index of refraction, as it is impossible to achieve agreement between $\fexp$ and $\ftheo$ in the whole frequency range by choosing a fixed value of $n$. This is also true for the other two parameters, the radius $R$ and the thickness $b$, or combinations thereof. Thus we can exclude the possibility of badly determined parameters and inaccuracies in the measurement of the resonance frequencies and state our main result: the $\neff$-model does not correctly describe the measured resonance frequencies. \newline
Finally, the measured and calculated resonance widths (FWHM) are compared. The experimental resonance widths $\Gexp$ are obtained by fitting Lorentzians to the measured spectrum. They consist of three terms,
\begin{equation} \Gexp = \Grad + \Gabs + \Gant , \end{equation}
where $\Grad$ describes the losses due to radiation, $\Gabs$ the losses due to absorption in the Teflon and $\Gant$ the loss of power due to the coupling to the antennas. The calculated resonance widths
\begin{equation} \Gtheo = -2 c \, \Im{k} / (2 \pi) \end{equation}
include only the losses due to radiation ($\Grad$). The measured and calculated widths of the TM-modes for disk A are shown in \reffig{fig:Gamma_HKreis}. For low frequencies (up to $8$ GHz for $n_r = 1$ and up to $10.5$ GHz for $n_r = 2$), the calculated widths are up to twice as large as the measured widths. Since $\Gtheo$ does not account for absorption and antenna losses, the actual difference is even larger. For higher frequencies, the measured widths are larger than the calculated ones and saturate at a value of about \mbox{$4$ MHz}. This saturation is due to absorption in the Teflon material and coupling to the antennas, and is approximately independent of the frequency. A precise comparison with the calculated widths is generally not possible because the radiation losses $\Grad$ cannot be extracted from the measured widths $\Gexp$. Nonetheless it can be stated that the widths $\Gtheo$ predicted by the $\neff$-model are too large at least in some frequency ranges within the range of accuracy of the index of refraction $n$. It should also be noted that $\Gtheo$ does not depend as sensitively on $n$ as the resonance frequencies. For TE-modes, the same general trend for $\Gexp$ and $\Gtheo$ is found, although the difference between $\Gexp$ and $\Gtheo$ is not as pronounced as for the TM-modes.

\section{\label{sec:NGKreis}Comparison of model and experiment for disk B}

\begin{figure}[tbh]
\subfigure[\,TE-modes]{
	\includegraphics[width = 8.4 cm]{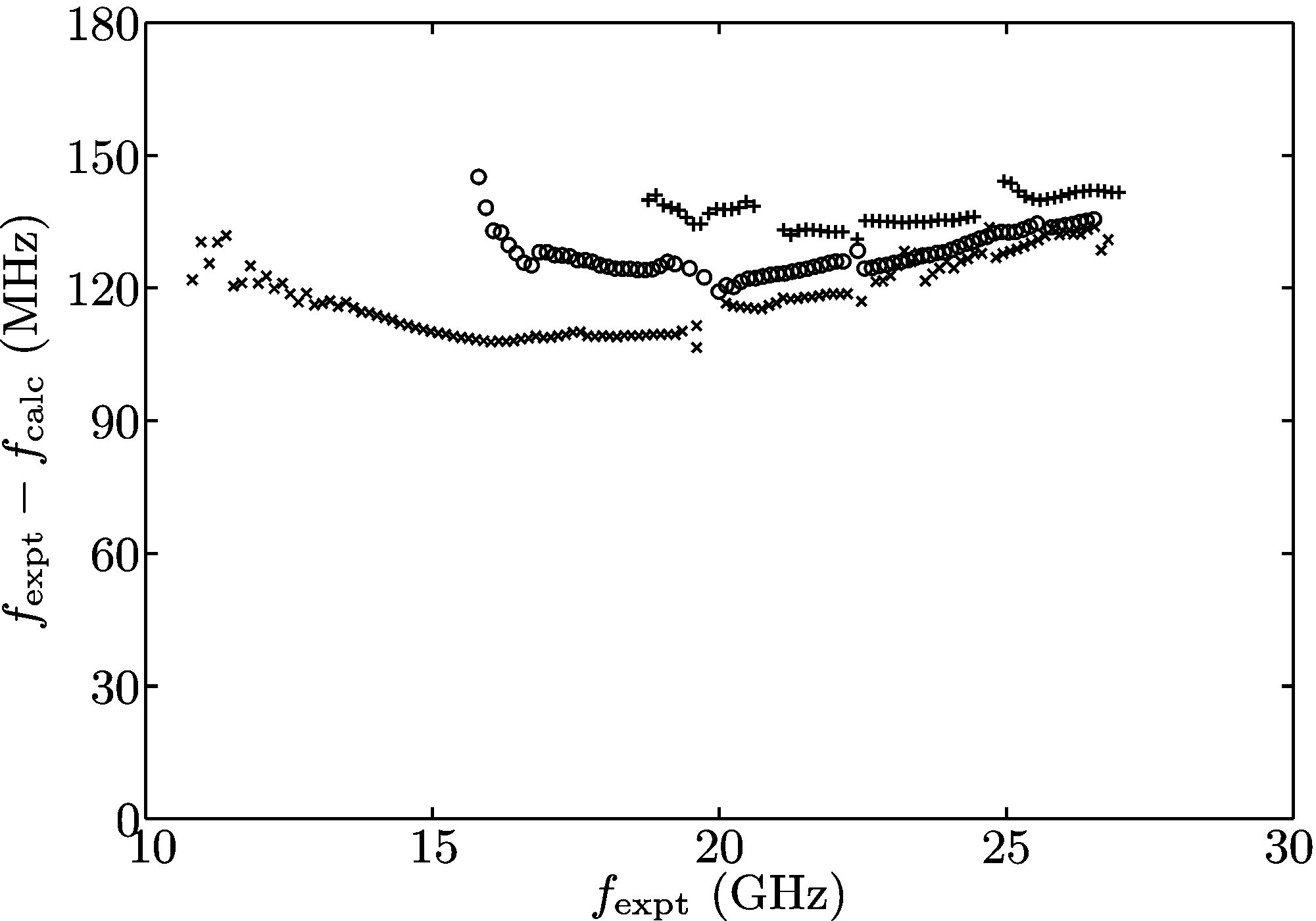}
	\label{subfig:fe-ft_NGKreis_TE}
}
\subfigure[\,TM-modes]{
	\includegraphics[width = 8.4 cm]{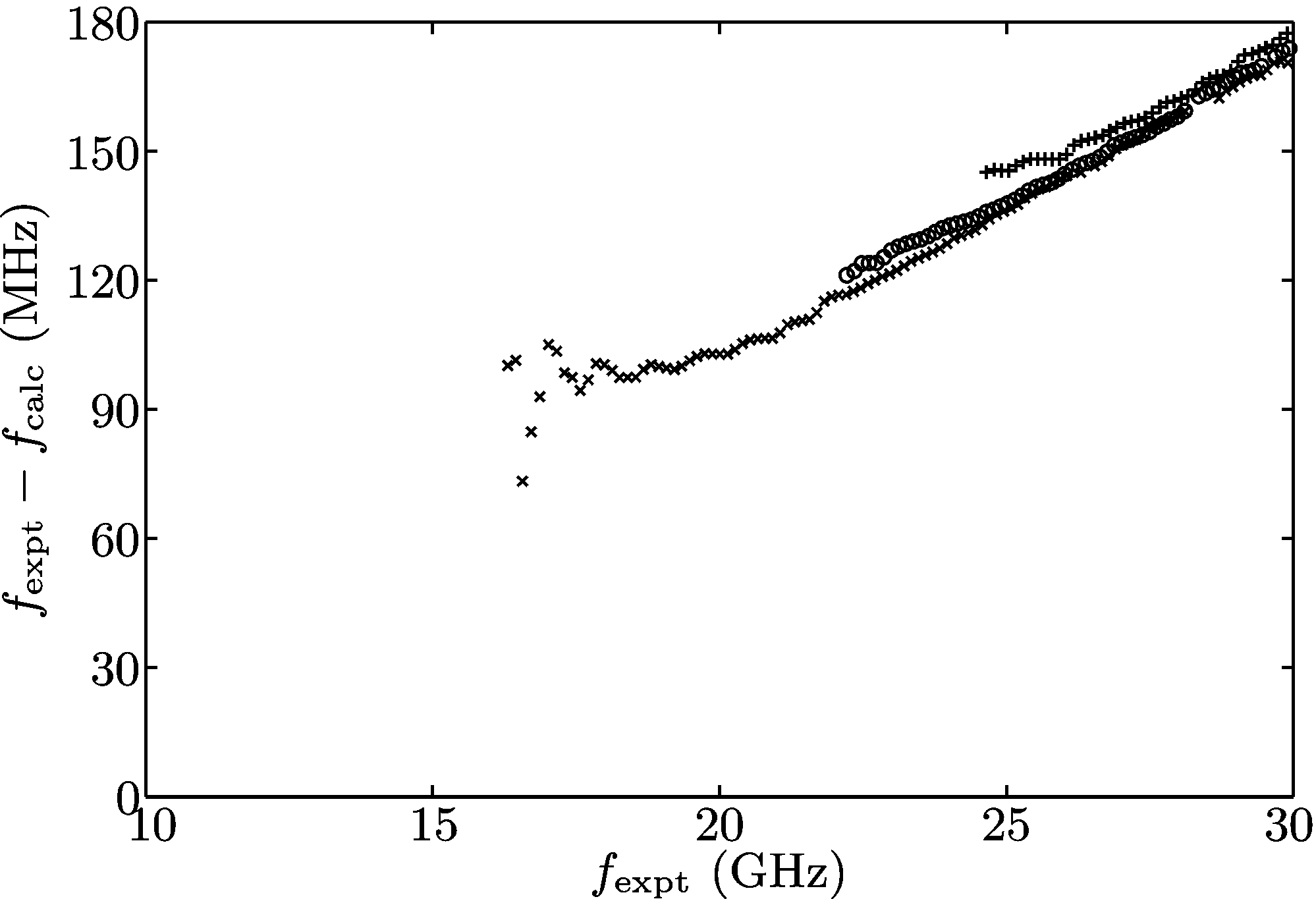}
	\label{subfig:fe-ft_NGKreis_TM}
}
\caption{\label{fig:fe-ft_NGKreis}Difference between measured ($\fexp$) and calculated ($\ftheo$) resonance frequencies with respect to $\fexp$ for disk B. The different symbols correspond to the different radial quantum numbers ($\times$: $n_r = 1$, $\circ$: $n_r = 2$, $+$: $n_r = 3$). The TE-modes measured with curved antennas are plotted in graph \subref{subfig:fe-ft_HKreis_TE} and have a range of azimuthal quantum numbers $m = 64$--$188$ for resonances with $n_r = 1$. The TM-modes shown in graph \subref{subfig:fe-ft_HKreis_TM} were measured with dipole antennas and have azimuthal quantum numbers $m = 97$--$204$ for $n_r = 1$. The frequency range of identifiable TE- and TM-modes differs due to the different quality factors and types of antennas used.}
\end{figure}

\begin{figure}[tbh]
\subfigure[\,TE-modes]{
	\includegraphics[width = 8.4 cm]{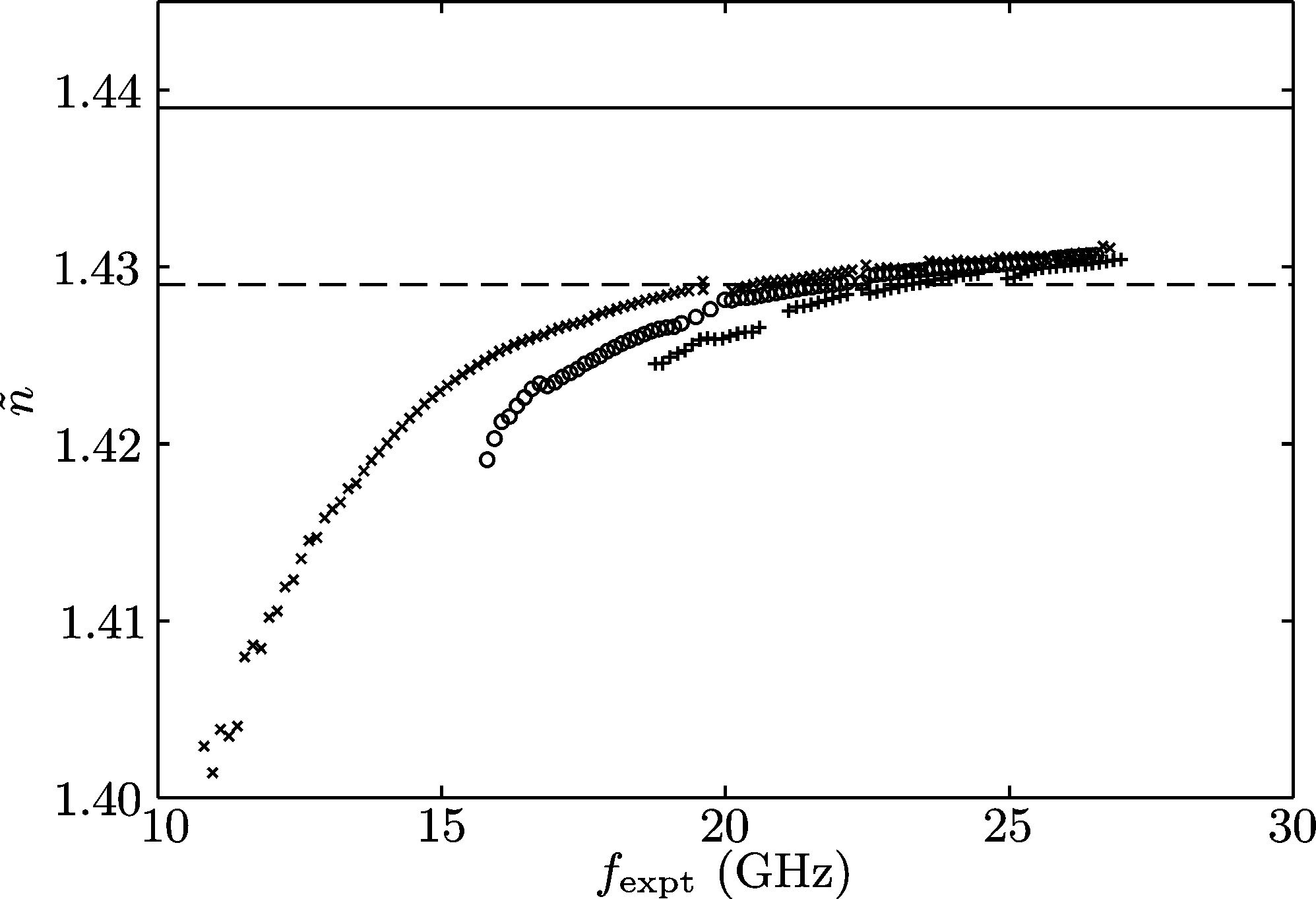}
	\label{subfig:nexp_NGKreis_TE}
}
\subfigure[\,TM-modes]{
	\includegraphics[width = 8.4 cm]{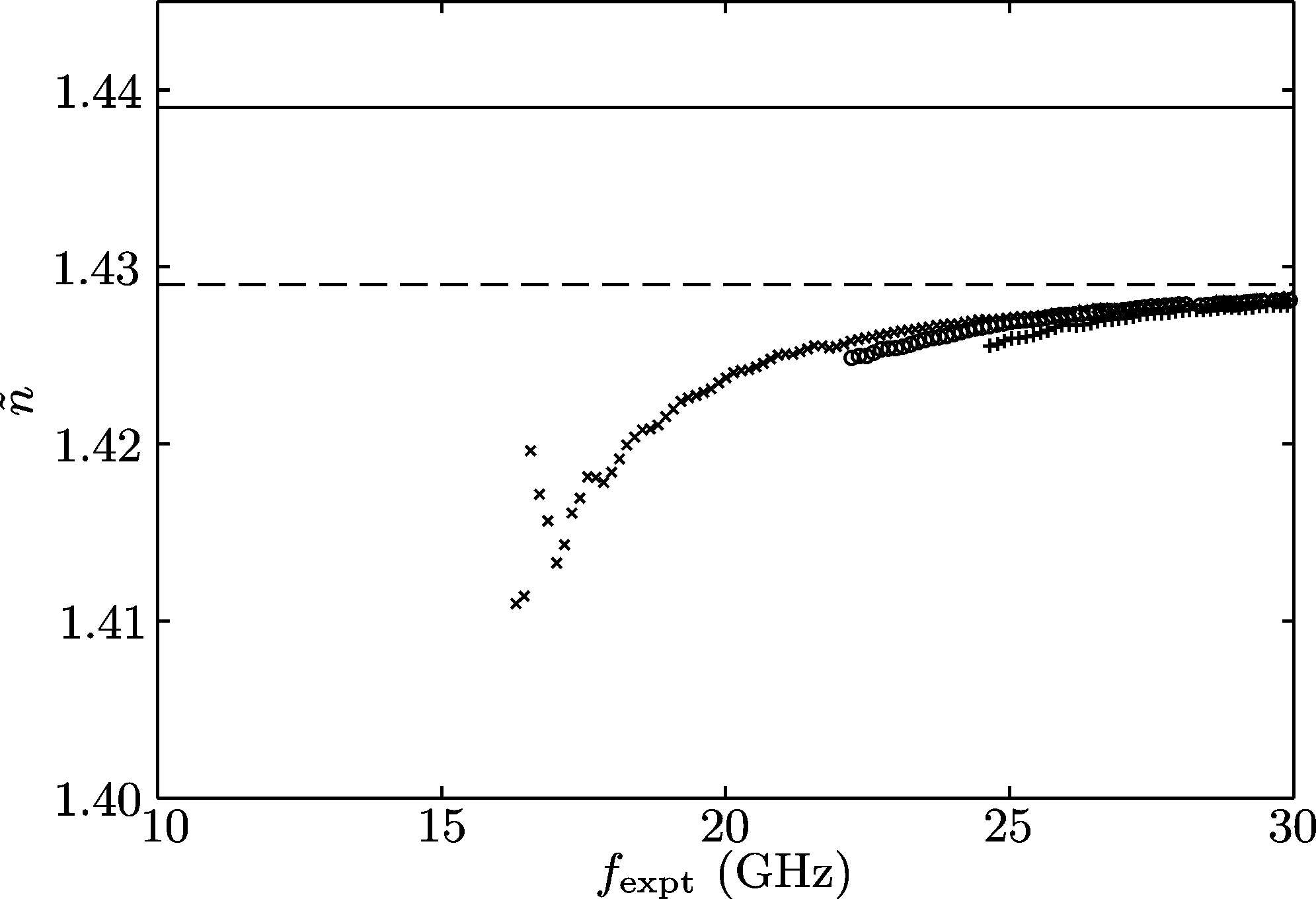}
	\label{subfig:nexp_NGKreis_TM}
}
\caption{\label{fig:nexp_NGKreis}Index of refraction $\nexp$ required to reproduce the measured resonance frequencies with \refeq{eq:qbed_diel_circ} as a function of the resonance frequency $\fexp$ for disk B. Each symbol corresponds to a radial quantum number ($\times$: $n_r = 1$, $\circ$: $n_r = 2$, $+$: $n_r = 3$), the solid line denotes the index of refraction $n = 1.439$ of the disk and the dashed line $n - \Delta n$ the range of accuracy of its determination. Part \subref{subfig:nexp_NGKreis_TE} shows the data points for the TE-modes and part \subref{subfig:nexp_NGKreis_TM} those for the TM-modes. The trend of the curves is comparable to \reffig{fig:nexp_HKreis}, but the deviations from $n$ are even larger.}
\end{figure}

\begin{figure}
\includegraphics[width = 8.6 cm]{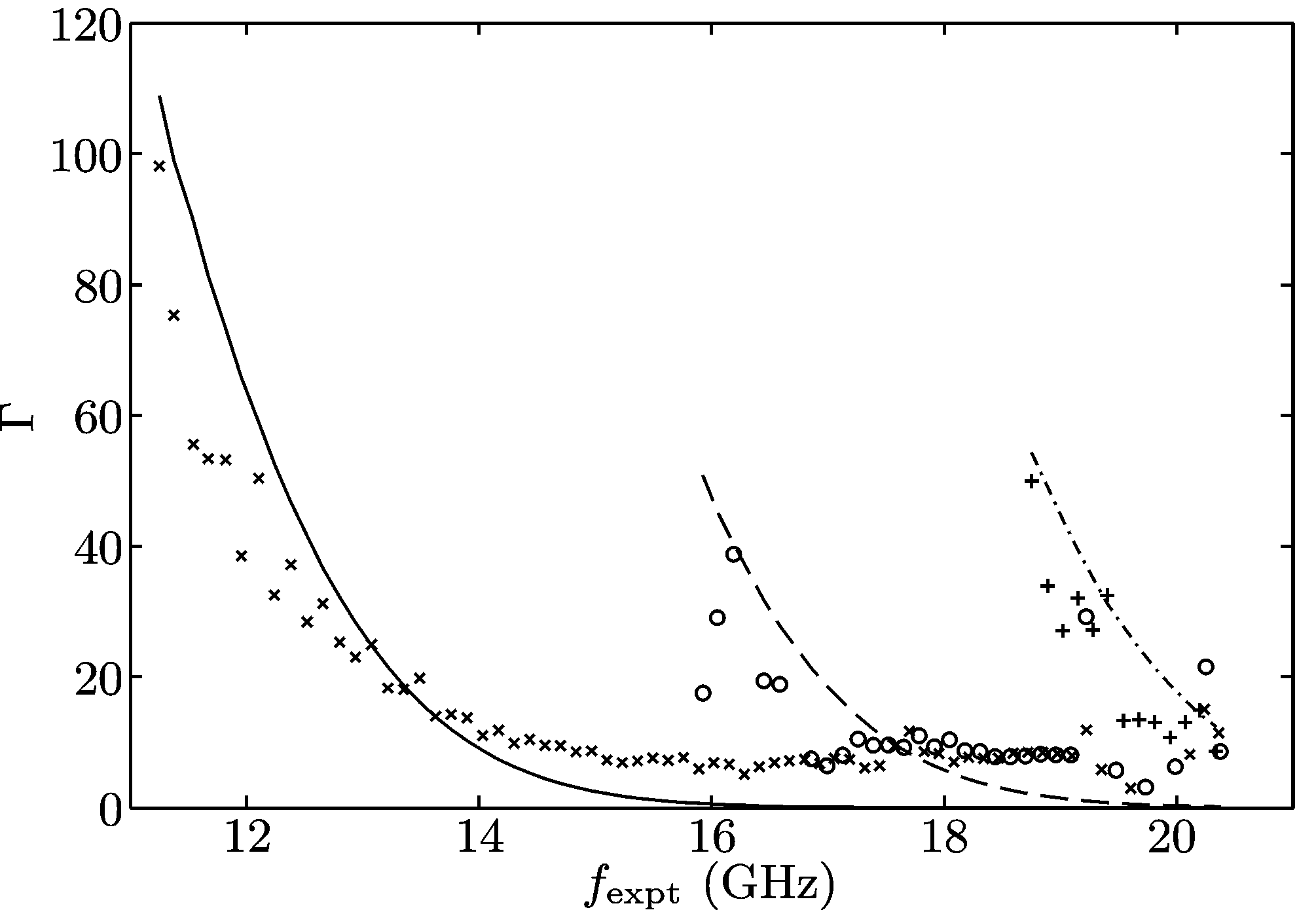}
\caption{\label{fig:Gamma_NGKreis}Measured ($\Gexp$) and calculated ($\Gtheo$) resonance widths for the TE-modes of disk B. Both $\Gexp$ and $\Gtheo$ are plotted with respect to the measured resonance frequency $\fexp$. The different symbols denote the measured widths for different radial quantum numbers ($\times$: $n_r = 1$, $\circ$: $n_r = 2$, $+$: $n_r = 3$). The calculated widths are plotted as curves (solid line for $n_r = 1$, dashed for $n_r = 2$ and dot-dashed for $n_r = 3$) instead of data points to guide the eye.}
\end{figure}

Next, we will compare the data of the second, thinner disk B (\mbox{$b = 5.0$ mm}) with the calculations based on the $\neff$-model. The discrepancy between experiment and model becomes apparent when comparing measured and calculated resonance frequencies in \reffig{fig:fe-ft_NGKreis} (like in \reffig{fig:fe-ft_HKreis} for disk A). The differences $\fexp - \ftheo$ seem to be larger for disk B, but are of the same order of magnitude. However, in contrast to those for disk A, they increase with increasing excitation frequency. In fact, this behavior depends sensitively on $n$; changing the value of $n$ within the range of accuracy leads to differences increasing, decreasing or reaching a finite value with increasing $\fexp$ (see also \reffig{fig:fe-ft_3n}). In any case the deviations are larger than for disk A at least for frequencies up to \mbox{$20$ GHz}. \newline
Like in \reffig{fig:nexp_HKreis}, the index of refraction $\nexp$ needed to reproduce the measured resonance frequencies with \refeq{eq:qbed_diel_circ} is shown for $n_r = 1, 2, 3$ in \reffig{fig:nexp_NGKreis} with respect to the resonance frequency $\fexp$. Qualitatively, the curves are similar to those in \reffig{fig:nexp_HKreis}: $\nexp$ grows with increasing frequency and seems to converge to some value depending on the polarization, and $\nexp$ has a weak dependence on the radial quantum number $n_r$ of the modes. Compared to the results for disk A shown in \reffig{fig:nexp_HKreis} the deviation of $\nexp$ from $n$ seems to be even larger for the thinner disk B, and those between the curves for different $n_r$ are more pronounced. Again, the systematic failure of the $\neff$-model is clearly visible, and the discrepancies seem to be even larger. \newline
The comparison of calculated and measured resonance widths for disk B is plotted in \reffig{fig:Gamma_NGKreis}. Like in \reffig{fig:Gamma_HKreis}, the measured widths $\Gexp$ are significantly smaller than the calculated widths $\Gtheo$ for the resonances with lower azimuthal quantum numbers, although the difference is not quite as large as for the case of disk A. With increasing frequency, the measured widths saturate at a value of about \mbox{$7$ MHz}, and a comparison with the calculated widths is not possible because of the additional losses due to absorption and the antennas. The difference between $\Gexp$ and $\Gtheo$ is smaller in the case of TM-modes, and gets smaller for both cases if a larger index of refraction $n$ is assumed in the calculations. Still, it is definite that the calculated widths are too large at least for some resonances.

\section{\label{sec:conc}Conclusions}
We have measured the resonance frequencies and widths of two different flat cylindrical dielectric microwave resonators. The quantum numbers and the polarization of the corresponding modes were identified. These data were used to test the $\neff$-model, a two-dimensional approximation for flat dielectric resonators with large planar extension normally used for the modeling of e.g.\ microlasers. A microwave resonator was chosen as testbed for several reasons: It is easy to handle due to its macroscopic dimensions which are known with high precision, the measurement of intensity distributions enables a detailed understanding of the spectrum and the identification of individual resonances, and it is a passive system which means that additional shifts of resonance frequencies and widths due to the lasing process in an active medium \cite{Harayama2005, Harayama2003b} can be excluded. \newline
It was shown that the $\neff$-model fails to correctly predict the resonance frequencies in a systematic way, and it has been checked carefully that the deviations between experiment and calculations are not due experimental inaccuracies caused by e.g.\ the antennas or the suspensions. Also, the resonance widths are clearly overestimated at least in the lower frequency range. This is important because the widths (of the passive cavity) determine the lasing threshold in microlasers and play a crucial role with regard to mode-competition \cite{Harayama2005, Harayama2003b}. Although the calculations based on the $\neff$-model yield at least the correct order of magnitude for both resonance frequencies and widths, a detailed understanding of the spectrum with help of these calculations is impossible. Furthermore, the accuracy of the model is not under control: The magnitude of the deviations between model and experiment cannot be precisely determined due to the uncertainty in the index of refraction, and depends on the dimensions of the cavity and the frequency in a non-trivial way. It seems that deviations are larger for smaller values of $kb$, i.e.\ thinner resonators. We believe that the main reason for the failure of the $\neff$-model is the incorrect formulation of the boundary conditions (as in \refeq{eq:neff_bcs}). In the rigorous treatment of the 3D Maxwell equations the $z$-components of the electric and magnetic fields are connected by the boundary conditions and cannot be considered separately as done in the $\neff$-model \cite{Schwefel2005, Dubertrand2008}. \newline
The large deviations between model and experiment observed for the circular disk are not due to the localization of the (whispering gallery) modes close to the boundary, and not expected to be smaller if modes supported by the whole area of the disk were considered. In fact the modes with lower $n_r$ are localized closer to the boundary (see \reffig{fig:WFs}), but the deviations are smaller than for modes with higher $n_r$. Even though the circular resonator is a special case and the test of the $\neff$-model with a non-circular geometry remains an open problem, it is still worthwhile to consider it: the quasi-bound states of many other non-circular resonators are also of whispering gallery mode type. Moreover, the circular resonator has many important applications by itself \cite{Annino1997, Annino2000}. One of these is the precise measurement of the index of refraction over a wide range of frequencies which, however, relies on a rigorous model for the resonance frequencies. \newline
In summary, the comparison of the measured and computed resonance frequencies and widths clearly attests the need of an improved $\neff$-model, which takes into account the boundary conditions at the cylindrical sidewalls. We hope that this work will stimulate further research in this direction to obtain a reliable 2D model for the computation of the resonance frequencies and widths of flat dielectric resonators. Although the numerical solution of the full three-dimensional Maxwell's equations is feasible for the circular resonator \cite{Vuckovic1999, Ghulinyan2008, Song2009}, it is computationally demanding, and even more so for other (viz.\ chaotic) geometries. Still, 3D numerical calculations for the circular resonator would be helpful to validate our data and investigate the accuracy of the $\neff$-model in a broader range of aspect ratios $b/R$ and indices of refraction $n$.

\begin{acknowledgments}
The authors are grateful to E. Bogomolny for his original suggestion to investigate the validity of the $\neff$-model experimentally and to him and M. Hentschel for many intense discussions. F. S. acknowledges support from Deutsche Telekom Foundation. This work was supported by the DFG within the Sonderforschungsbereich 634.
\end{acknowledgments}



\end{document}